\documentclass[pdflatex,sn-mathphys-num]{sn-jnl}

\usepackage{graphicx}%
\usepackage{multirow}%
\usepackage{amsmath,amssymb,amsfonts}%
\usepackage{amsthm}%
\usepackage{mathrsfs}%
\usepackage[title]{appendix}%
\usepackage{xcolor}%
\usepackage{textcomp}%
\usepackage{manyfoot}%
\usepackage{booktabs}%
\usepackage{algorithm}%
\usepackage{algorithmicx}%
\usepackage{algpseudocode}%
\usepackage{listings}%
\usepackage{graphicx,amsmath,amssymb} 
\usepackage{algorithm,algpseudocode}
\usepackage{subcaption} 
\usepackage{color} 
\usepackage{blindtext}
\usepackage{hyperref}
\usepackage{amsthm}
\usepackage{placeins} 

\theoremstyle{thmstyleone}%
\theoremstyle{thmstyletwo}%
\newtheorem{remark}{Remark}%

\theoremstyle{thmstylethree}%

\begin{document}

\title[Neural-POD]{Neural-POD: A Plug-and-Play Neural Operator Framework for Infinite-Dimensional Functional Nonlinear Proper Orthogonal Decomposition}

\author[1]{\fnm{Changhong} \sur{Mou}} \email{changhong.mou@usu.edu}

\author[2]{\fnm{Binghang} \sur{Lu}} \email{lu895@purdue.edu}

\author*[3,4]{\fnm{Guang} \sur{Lin}}\email{guanglin@purdue.edu}

\affil[1]{\orgdiv{Department of Mathematics and Statistics},
           \orgname{Utah State University},
           \orgaddress{\street{3900 Old Main Hill},
                       \city{Logan},
                       \state{UT},
                       \postcode{84322},
                       \country{USA}}}

\affil[2]{\orgdiv{School of Electrical and Computer Engineering},
           \orgname{Purdue University},
           \orgaddress{\street{610 Purdue Mall},
                       \city{West Lafayette},
                       \state{IN},
                       \postcode{47907},
                       \country{USA}}}

\affil[3]{\orgdiv{Department of Mathematics},
           \orgname{Purdue University},
           \orgaddress{\street{610 Purdue Mall},
                       \city{West Lafayette},
                       \state{IN},
                       \postcode{47907},
                       \country{USA}}}

\affil[4]{\orgdiv{School of Mechanical Engineering},
           \orgname{Purdue University},
           \orgaddress{\street{610 Purdue Mall},
                       \city{West Lafayette},
                       \state{IN},
                       \postcode{47907},
                       \country{USA}}}


\abstract{
AI for science (AI4Science) models often suffer from discretization: learned representations remain tied to the training grid, limiting transfer across resolutions, solvers and applications. We introduce Neural Proper Orthogonal Decomposition (Neural-POD), a plug-and-play neural operator that learns nonlinear, orthogonal basis functions directly in function space and can be integrated in both projection-based reduced order models and operator-learning frameworks such as DeepONet. Neural-POD replaces SVD-derived, resolution-dependent linear modes with continuous, resolution-invariant bases learned via sequential residual minimization, analogous to Gram–Schmidt orthogonalization. The framework supports training under task-specific norms (e.g., $L^2$, $L^1$), improves out-of-distribution generalization to unseen parameter regimes, and captures nonlinear structure in complex systems. Because the learned bases are interpretable and reusable, Neural-POD serves as a general representation module for AI4Science workflows. We demonstrate Neural-POD on Burgers’ and Navier–Stokes equations.}
\par\vspace{-1.2\baselineskip}



\maketitle

AI for science (AI4Science) increasingly relies on learned representations that must transfer reliably across discretizations, geometries, and parameter regimes, rather than being tied to a particular grid or snapshot resolution. In this setting, a central challenge is to build low-dimensional structures that are both faithful to the underlying infinite-dimensional dynamics and reusable across solvers and data that supports efficient simulation without retraining from scratch whenever the discretization changes. 
Consequently, Proper Orthogonal Decomposition (POD) is widely used in AI4Science as a physics-informed representation tool that provides compact low-dimensional bases for high-dimensional data that support efficient learning, inference and prediction across different scientific tasks.
POD, also known as principal component analysis (PCA), is a foundational technique in computational modeling for extracting dominant coherent structures from high-dimensional data. 
Historically, it emerged from early ideas in statistics and stochastic-process theory (PCA and the Karhunen--Lo\`eve viewpoint) and was later adopted in fluid mechanics and turbulence as a dominant way to identify energetic coherent modes from spatiotemporal measurements and simulations \cite{pearson1901liii,hotelling1933analysis,lumley1967similarity}. 
POD supports a broad range of methodologies, including reduced order modeling (ROM), where POD modes provide low-rank representations that enable efficient simulation while preserving essential dynamics \cite{benner2015survey,lucia2004reduced,yu2019non}, and scientific machine learning (SciML), where POD-based decompositions offer structured low-dimensional representations for learning operators and dynamics (for example, POD--DeepONet architectures \cite{lu2022comprehensive}). 
Beyond ROM and operator learning, POD is widely used for data compression, feature extraction and modal analysis in complex spatiotemporal systems. 
Classical POD is typically computed via singular value decomposition (SVD), yielding $L^2$-optimal linear basis functions from snapshot data \cite{HLB96}. 
Yet, when modern applications demand robustness for different discretizations, diverse regimes and strongly nonlinear behaviours, traditional POD faces persistent obstacles.

These obstacles are well recognized. 
First, POD is often resolution-dependent: modes learned on one discretization may lose optimality or even effectiveness when the grid changes, limiting transferability across resolutions \cite{taira2017modal}. 
Second, POD may lose accuracy even for in-distribution interpolation and can fail more severely for out-of-distribution extrapolation beyond the training regime, which limits its reliability in previously unseen scenarios \cite{carlberg2011efficient}.
Third, although POD is optimal in an $L^2$ sense within the training dataset, its linear subspace structure can be too restrictive to represent the nonlinear features that drive many physical systems \cite{quarteroni2015reduced, benner2015survey}. 
As a consequence, sharp gradients, discontinuities and other strongly nonlinear structures may be poorly captured, especially when such features dominate the dynamics \cite{rowley2004model}. 
Taken together, these challenges complicate deployment in multi-resolution regimes and hinder real-time simulation, where one needs a compact representation that generalizes reliably as dynamical or physical regimes change.

To address these limitations, we propose \textbf{\textit{neural proper orthogonal decomposition}} (Neural-POD), a drop-in framework that replaces linear POD modes with \emph{nonlinear} basis functions parameterized by neural networks.
The key idea is to preserve the progressive, mode-by-mode error reduction of POD while upgrading the representational power and enabling transfer across resolutions and regimes.
Given snapshot data, Neural-POD learns the first mode by minimizing a reconstruction loss between the snapshots and a neural representation with a learnable time-dependent coefficient; subsequent modes are learned sequentially by training new networks on the residual from the previous approximation \cite{shadden2011lagrangian,yosinski2014transferable,deng2021deeplight}. 
This iterative procedure yields an orthogonal set of Neural-POD modes while directly retaining the structures most relevant to the data.
Crucially, after training, \emph{only the Neural-POD model parameters} (and the associated low-dimensional coefficients) need to be stored, not resolution-specific basis vectors nor the full snapshot matrix, which makes the representation compact and easy to deploy.
This compactness enables rapid online evaluation and supports real-time or many-query tasks, where one repeatedly updates reduced states or operators under limited computational budgets.

Neural-POD offers additional benefits that directly target the above challenges. 
First, the training objective is flexible that allows optimization in task-relevant norms (for example $L^2$, $L^1$ or $H^1$), rather than being prescribed to a fixed $L^2$ criterion. 
Second, representing modes as neural functions admits resolution independence and improves in-distribution and out-of-distribution across multi-resolution settings \cite{mendez2019multi}. 
Third, the learned representation can be reused across parameter variations with substantially reduced retraining burden, enabling efficient parametric studies. 
Finally, Neural-POD can serve as a pretrained, drop-in component inside broader operator-learning frameworks \cite{zahr2015progressive} that provides a structured latent representation that is both compact to store and fast to evaluate for real-time development.

\begin{figure}[t]
    \centering
    \begin{subfigure}[t]{\linewidth}
        \centering
        \includegraphics[width=0.8\linewidth]{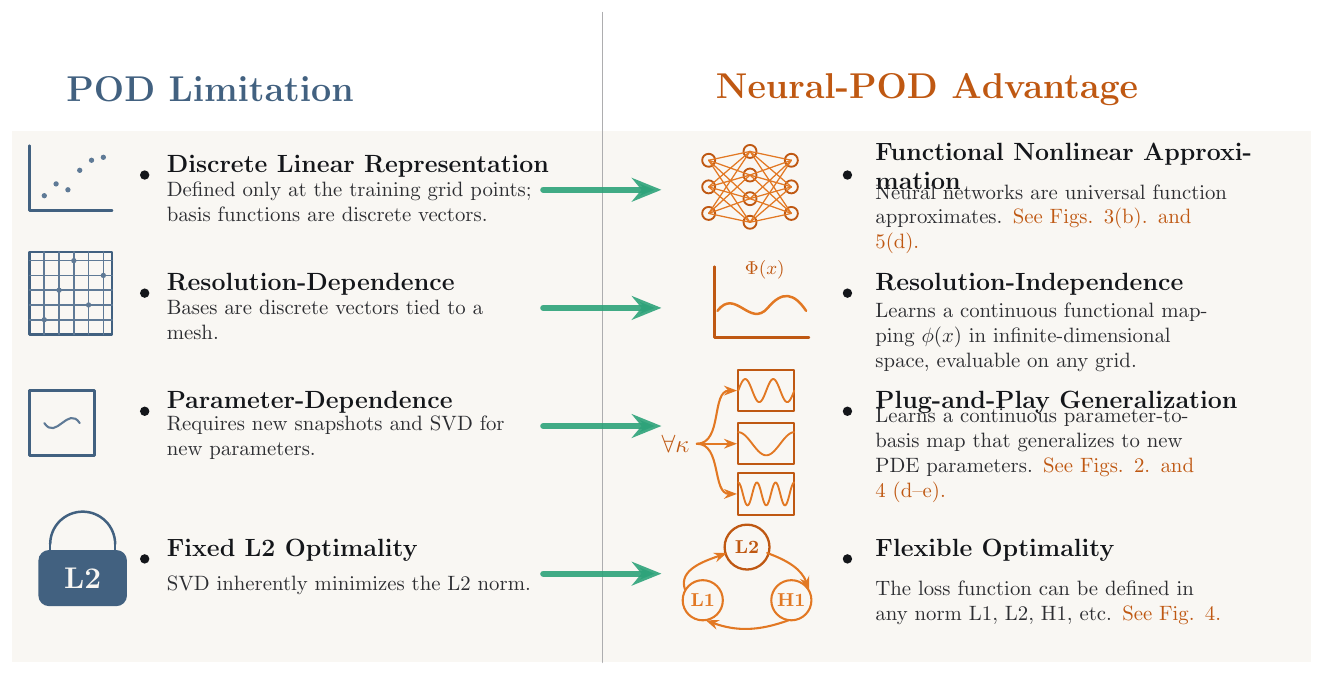}
        \caption{
        Comparison between classical POD and Neural-POD.
        Classical POD constructs discrete, resolution-dependent basis vectors tied to training grids and parameters,
        whereas Neural-POD learns continuous, nonlinear, resolution-invariant basis functions expressed by neural networks,
        enabling parameter generalization and flexible optimality norms. \label{fig:neural_vs_pod}
        }
    \end{subfigure}
    \hfill
    \begin{subfigure}[t]{\linewidth}
        \centering
        \includegraphics[width=0.8\linewidth]{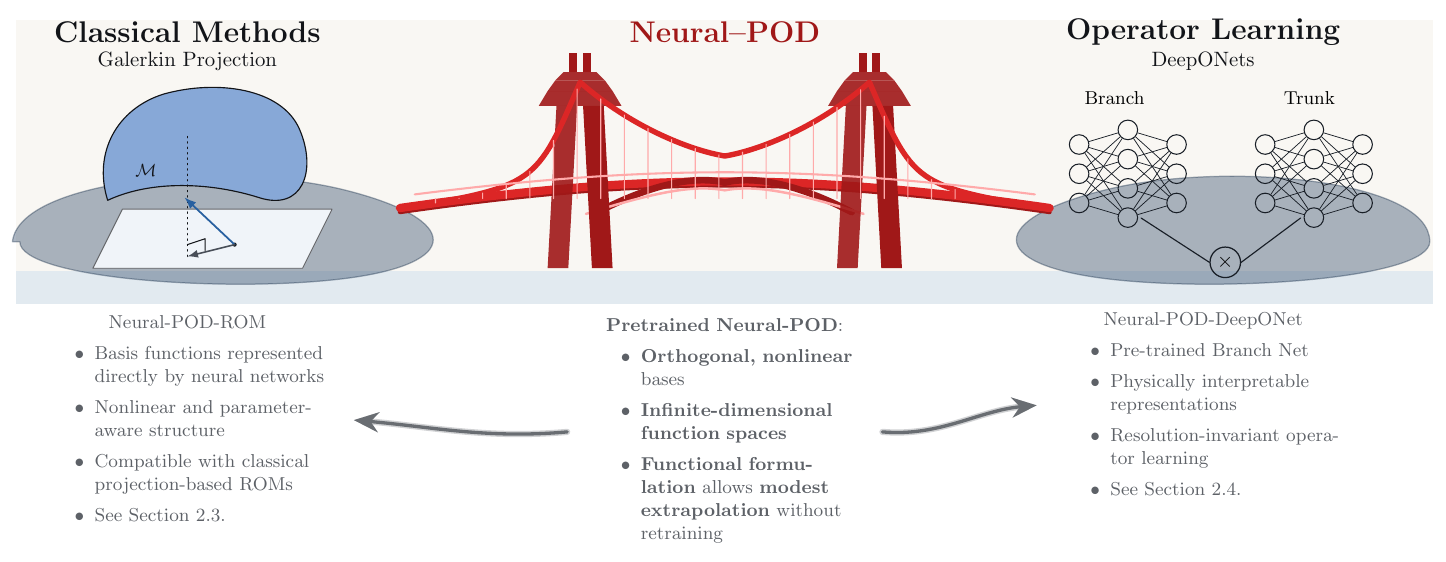}
        \caption{
        Neural-POD as a bridge between projection-based reduced-order models and operator learning.
        Neural-POD provides orthogonal, nonlinear basis functions for Galerkin ROMs while simultaneously serving as
        a pretrained, physically interpretable Branch Network within DeepONet architectures. 
                \label{fig:bridge_neural_pod}
        }
    \end{subfigure}
    \begin{subfigure}[t]{\linewidth}
        \centering
        \includegraphics[width=.8\linewidth]{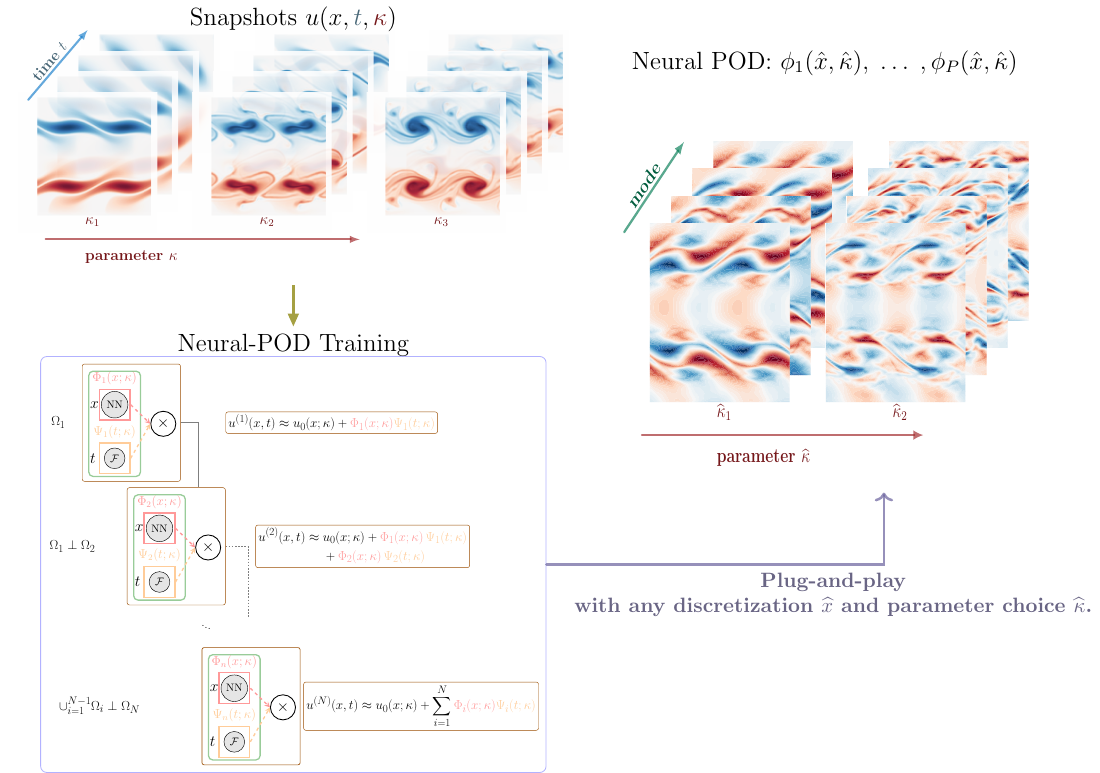}
        \caption{
            Overview of Neural-POD training. Snapshots $u(x,t,\kappa)$ are collected over time for varying parameter values $\kappa$ (upper left). The training architecture decomposes each snapshot into products of spatial mode functions $\Phi_i(x;\kappa)$ and temporal coefficients $\Psi_i(t;\kappa)$, which guarantees the orthogonality for all modes (bottom left). Once trained, the Neural-POD basis provides plug-and-play reduced representations that can be evaluated on any discretization $\hat{x}$ and parameter choice $\hat{\kappa}$ (right, $\hat{\kappa}$ may differ from the training snapshots’ parameter $\kappa$).
    \label{fig:npod-training}
        }
    \end{subfigure}
    \caption{
    Neural-POD for resolution-independent model reduction and operator learning.    }
    \label{fig:neural_pod_combined}
\end{figure}

\begin{figure*}[p]
    \centering

    \begin{subfigure}[t]{\linewidth}
        \centering
        \includegraphics[width=\linewidth]{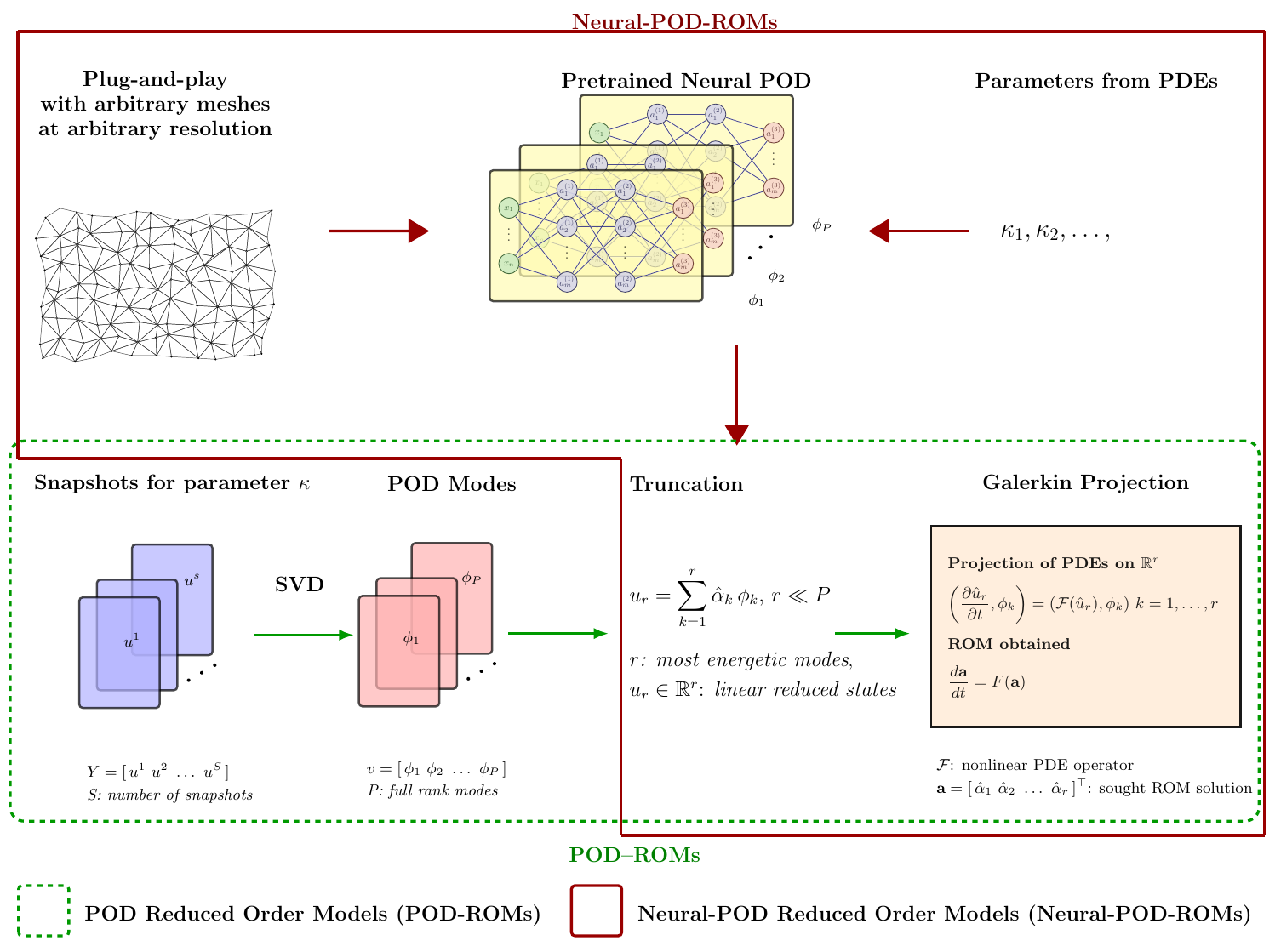}
        \caption{Schematic diagram of the proposed Neural-POD-ROMs and traditional POD-based reduced order models (POD-ROMs).  Top portion: the Neural-POD extension that enables plug-and-play operation on arbitrary meshes and at arbitrary resolution, informed by pretrained neural representations of POD bases and PDE parameters. Bottom portion: the traditional POD flowchart, including snapshot generation for parameter $\kappa$, computation of full POD modes via SVD, construction of the reduced basis, and Galerkin projection of the governing PDE.\label{fig:npod-rom}
        }
    \end{subfigure}

    \vspace{0.8em}

    \begin{subfigure}[t]{\linewidth}
        \centering
        \includegraphics[width=\linewidth]{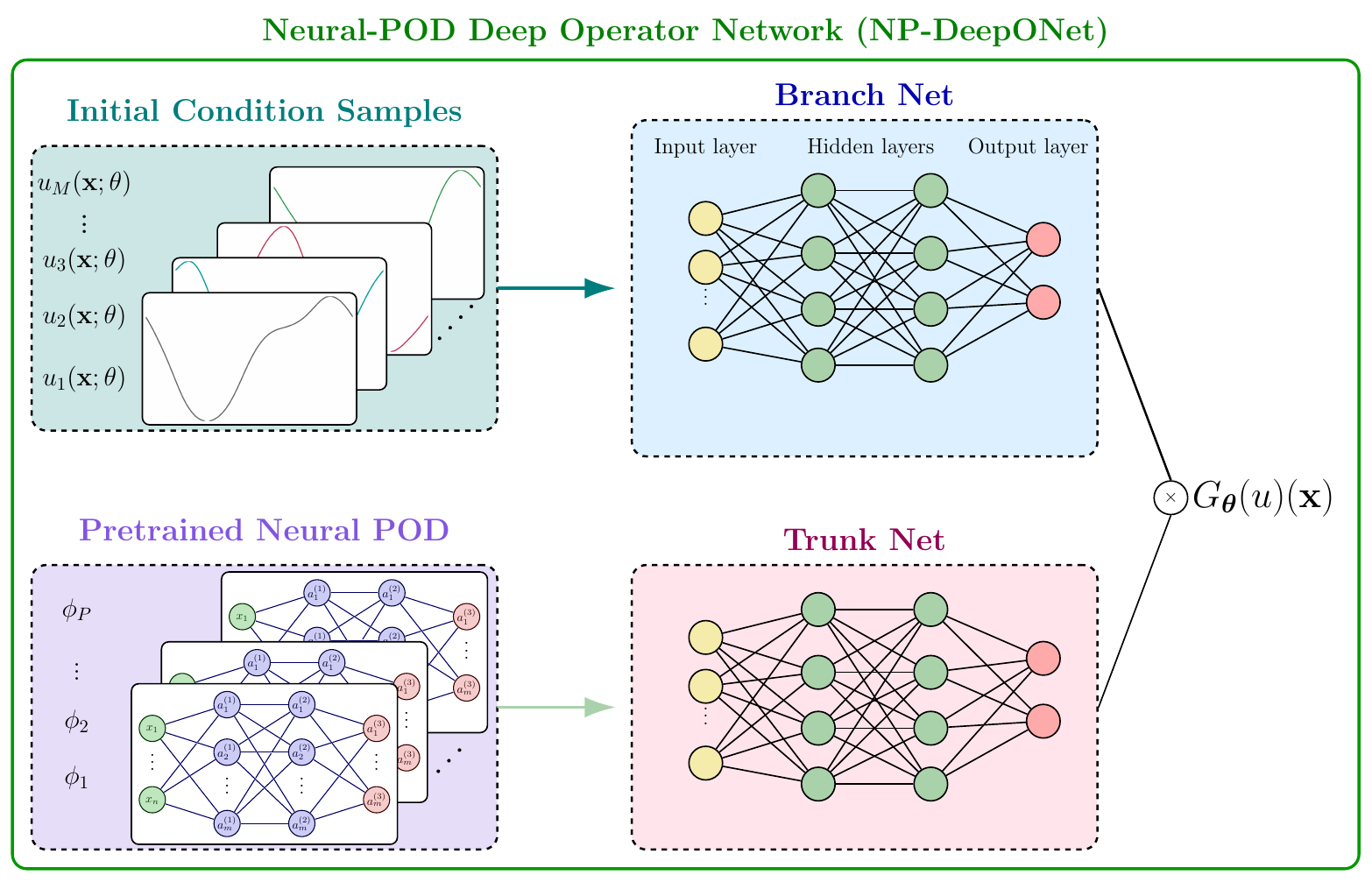}
        \caption{Schematic of the proposed Neural-POD Deep Operator Network (NP-DeepONet). The branch network takes initial condition fields as input, while the trunk network evaluates functions at spatial query points. A pretrained Neural-POD module provides reduced basis information ${\phi_1, \phi_2, \ldots, \phi_P}$ obtained from snapshot data. The outputs are combined to produce the operator evaluation $G_{\theta}(u_m)(x)$, enabling efficient and data-driven surrogate modeling of PDE solutions in reduced coordinates.         \label{fig:npod-deeponet}
}
    \end{subfigure}

    \caption{Neural-POD integrations for reduced order modeling and operator learning.}
    \label{fig:npod-rom-deeponet}
\end{figure*}

These properties position Neural-POD as a bridge can between projection-based ROM and modern deep learning.
In classical ROM, Neural-POD can be inserted into Galerkin projection frameworks to provide a more expressive and transferable basis while retaining the interpretability and structure of projection methods \cite{KV01,iliescu2014are,mou2021data}. 
In operator learning, Neural-POD offers a principle low-dimensional representation that can improve both efficiency and interpretability. 
For instance, Deep Operator Networks (DeepONets) comprise a branch network that encodes the input function sampled at sensor locations and a trunk network that encodes query locations for the output \cite{lu2019deeponet,winovich2025active,lu2025evolutionary}. 
By using Neural-POD as the branch network, DeepONets can incorporate orthogonal, data-driven nonlinear modes that capture underlying dynamics while remaining resolution-independent. 
This integration can improve generalization across discretizations and reduce training costs (since only Neural-POD can be pretrained), thereby enabling more efficient and robust operator learning in real-time or near-real-time settings.
In brief summary, our primary contributions include the following:

\begin{itemize}



\item \textbf{Neural-POD is a plug-and-play component} for both Galerkin ROM and operator learning: once the basis/model is learned offline, it can be dropped into a reduced order model with minimal modification, and the online stage only requires assembling and evolving the reduced system for fast, easy deployment across parameterized PDEs (see Figure~\ref{fig:bridge_neural_pod}). 
\textbf{This bridges traditional ROM and surrogate modeling with operator learning} by fitting naturally into familiar offline/online workflows and existing Galerkin solvers (see Figure~\ref{fig:npod-rom}).
Moreover, the same plug-and-play interface enables straightforward reuse across PDEs with different parameters which supports a pretrained, open-source database and a shared repository of benchmark problems spanning diverse geometries and parameter settings.

\item \textbf{Neural-POD also has strong educational value as an easy-to-use ROM and operator-learning tool.} (see Figure~\ref{fig:npod-deeponet}) Its modular, plug-and-play property makes it well-suited for classroom adoption (e.g., undergraduate numerical analysis, scientific computing, and introductory SciML courses), where students can rapidly learn reduced order models and operator learning, explore the offline/online procedures, and build intuition for projection-based modeling without heavy software overhead. This positions Neural-POD as a practical foundation for course modules, hands-on assignments, and open teaching resources with broad educational impact.


\item \textbf{Neural-POD enables accurate in-distribution prediction and reasonable generalization for out of distribution prediction} by learning resolution-independent basis functions that remain effective under modest parameter shifts, without retraining (see Figure~\ref{fig:neural-pod-interp}--\ref{fig:neural-pod-extrap} and Table \ref{tab:podrom-combined}).

\item \textbf{Neural-POD constructs nonlinear, orthogonal basis functions} through an iterative residual minimization procedure using with neural networks that is analogous to Gram--Schmidt orthogonalization and capable of capturing complex features (see Figure~\ref{fig:npod-training}).

\item \textbf{Neural-POD supports training under different norms} (e.g., $L^2$, $L^1$), which enables the learned bases to reflect smoothness, discontinuities, or other structural characteristics induced by the chosen norm (see Figures~\ref{fig:res-POD}--\ref{fig:pod-res-l1-12-b}).

\item \textbf{Neural-POD is formulated in infinite-dimensional function spaces}, which matters because it learns basis \emph{functions} rather than {grid-dependent vectors}. Therefore, the same modes can be evaluated consistently on different discretizations, enabling resolution-independent basis construction and seamless deployment across spatial resolutions (See Algorithm \ref{alg:neural-pod}).

\end{itemize}
The above novelties are also illustrated in Figures~\ref{fig:neural_vs_pod} and~\ref{fig:bridge_neural_pod}.

Rapid progress in AI4Science for reduced representations and operator learning has exposed a key gap: performance can be highly sensitive to discretization choices (resolution, numerical schemes, and discretization error), limiting reliability and generalization for different regimes. We address this by studying the robustness of Neural-POD’s nonlinear functional representations and integrating them into (i) projection-based ROMs and (ii) DeepONet-style operator learning. We further assess performance under a fixed training budget to identify when Neural-POD remains effective. Using Burgers’ equation and 2D Navier–Stokes, we benchmark Neural-POD against classical POD in reconstruction, predictive accuracy, and robustness to in-distribution and out-of-distribution parameters.
\raggedbottom

\FloatBarrier  
\clearpage
\thispagestyle{empty}
\clearpage

\FloatBarrier  
\clearpage
\thispagestyle{empty}
\clearpage


\section*{Results \label{sec:neural-pod-numeric}}

\subsection*{Neural-POD for Burgers' equation}
To demonstrate the capability of the proposed framework in addressing nonlinear PDEs, we use the one-dimensional (1D) Burgers' equation with periodic boundary conditions as discussed in Li \textit{et al.} \cite{li2020fourier,wang2021learning}.
\begin{align}
&\frac{\partial u}{\partial t} + u\frac{\partial u}{\partial x} - \nu \frac{\partial^2 u}{\partial x^2} = 0, 
&& (x,t) \in (0,1) \times (0,1],  \label{eq:burgers-a}\\
&u(x,0) = u_0(x), 
&& x \in (0,1),  \label{eq:burgers-b}\\
&u(0,t) = u(1,t), \quad 
\frac{du}{dx}(0,t)= \frac{du}{dx}(1,t), 
&& t \in (0,1).  \label{eq:burgers-c}
\end{align}
where $t \in (0,1)$, and the initial condition $u(x)$ is generated from one distribution of GRF $\sim \mathcal{N}(0,25^2(-\Delta + 5^2I)^{-4})$, satisfying the periodic boundary conditions.

\begin{figure}[H]
    \centering

    \begin{subfigure}[t]{\linewidth}
        \centering
        \includegraphics[width=\linewidth]{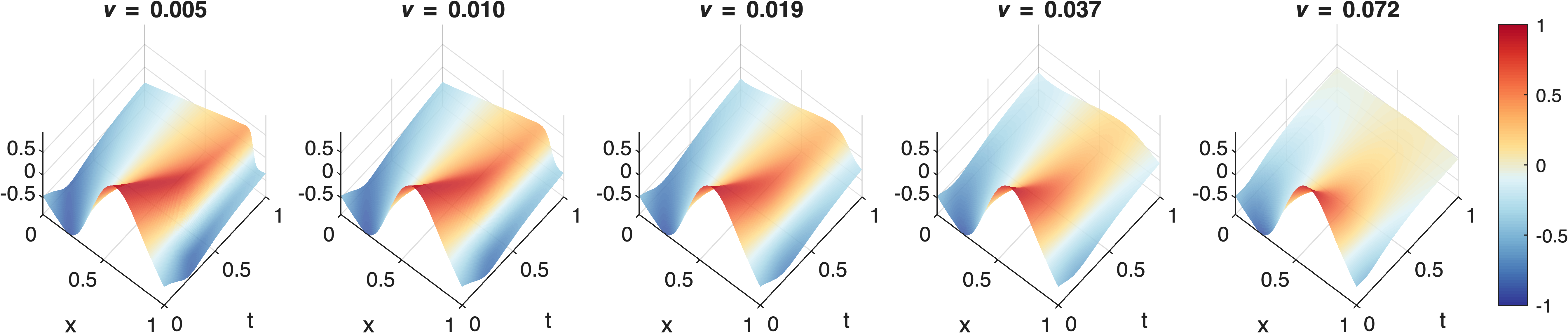}
        \caption{Solution snapshots of the Burgers equation at different sampled viscosity values $\nu$ used for Neural-POD training.}
        \label{fig:neural-pod-1}
    \end{subfigure}

    \vspace{0.8em}

    \begin{subfigure}[t]{\linewidth}
        \centering
        \includegraphics[width=\linewidth]{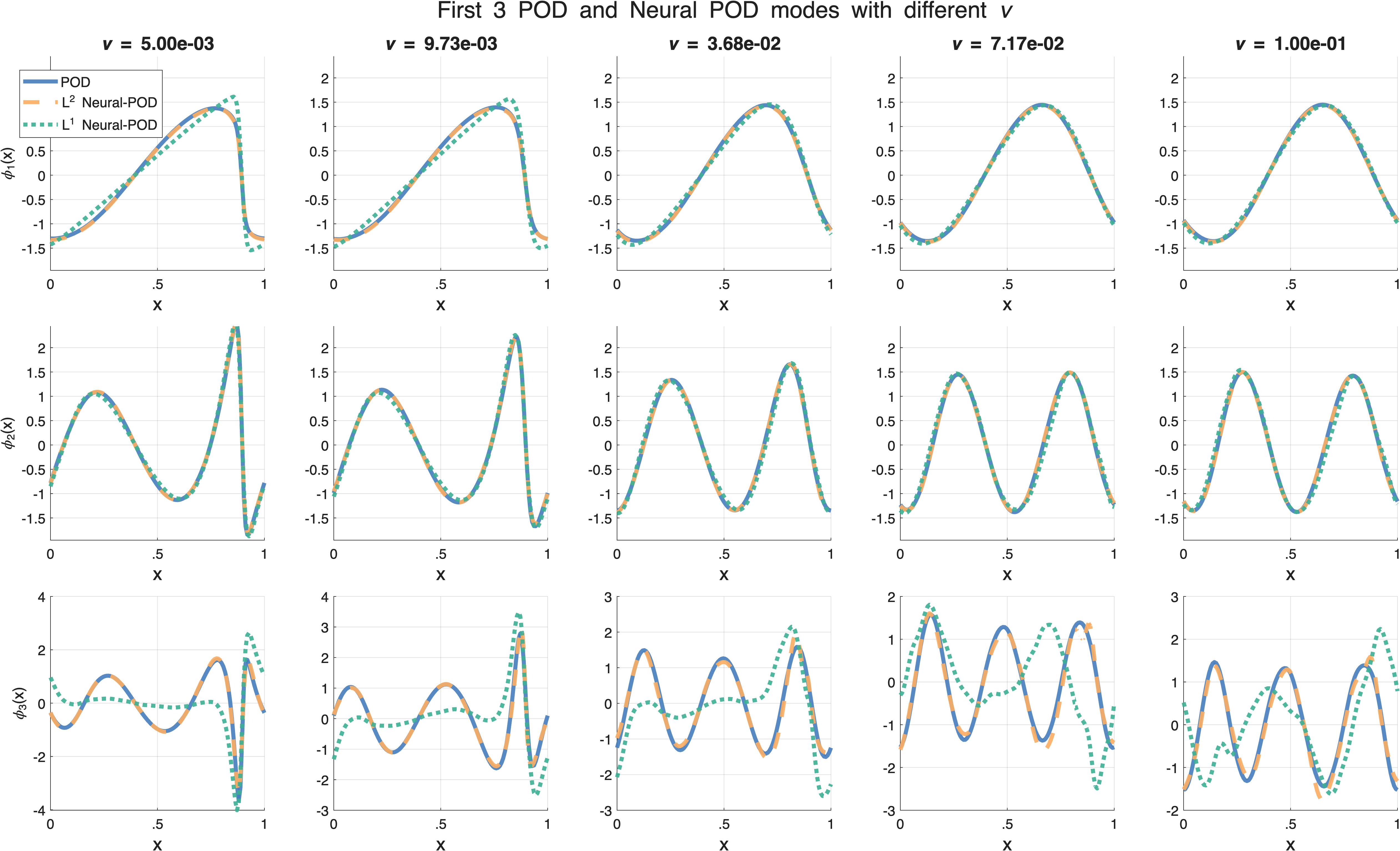}
        \caption{
        First three traditional POD modes and Neural-POD under different loss choices, i.e., $L^1$ and $L^2$ norms, for different viscosity values.
        }
        \label{fig:neural-POD}
    \end{subfigure}

    \caption{Burgers equations: (a) solution snapshots for different viscosities $\nu$; (b) traditional POD modes versus Neural-POD bases learned with $L^1$ and $L^2$ losses.}
    \label{fig:burgers-data-and-bases}
\end{figure}

In the following experiments, the Neural-POD model is trained using snapshot data generated at viscosity values $\nu$ sampled within the interval $[0.005,,0.01]$. This parameter range defines the training domain for the Neural-POD basis learning. 
Figure~\ref{fig:neural-pod-1} shows the evolution of the one-dimensional Burgers equation for different sampled viscosity values $\nu$ used to construct the training dataset for Neural-POD. 
As $\nu$ decreases (right to left), the solution transitions from a smooth diffusive profile to one dominated by advective nonlinearities, leading to the formation of sharp fronts and nearly discontinuous shock structures.
This coupled effects of convection and diffusion underscores the limits of traditional POD in handling unseen parameters, motivating parametric POD formulations that adapt to viscosity and smoothness variations. Extended Data Figure \ref{fig:traditional-POD}  shows the first three Neural-POD modes trained from FOM data for different viscosity values.
As the viscosity changes, the Neural-POD modes behaves significantly different behavior, with $\nu$ decreases, Neural-POD becomes increasingly oscillatory and sharply localized near the shock region.
Such sensitivity of the Neural-POD basis to the viscosity parameter indicates that the modes are not easily transferable across regimes, making it challenging to construct a compact parametric traditional POD representation that remains accurate for a wide range of different viscosity flows.


\paragraph{Neural-POD with $L^1$ and $L^2$ loss}
In addition to the standard $L^2$ based training loss, we also consider an $L^1$ norm formulation to show the robustness of the Neural-POD framework. The corresponding loss terms are defined as
\begin{align}
    &L^{1}\text{ Optimality:} &&L_{\text{data}} = \sum_{i=1}^N \left| \Phi(x_i)\Psi(t_i) - u_i \right|, \\
    &\text{Boundary conditions:} &&L_{\text{bc}} = \sum_{i=1}^N \left| \Phi(x^{bc}_i) - u^{bc}_i \right|^2 
    + \sum_{i=1}^N \left| 
    \frac{\partial \Phi(x^{bc}_i)}{\partial x} 
    - \frac{\partial u^{bc}_i}{\partial x} 
    \right|^2.
\end{align}
Compared to the $L^2$-norm, the $L^1$-based loss penalizes large residuals less aggressively, thereby reducing the influence of outliers and improving the representation of localized or discontinuous features in the data. 

\begin{figure}[t]
    \centering

    \begin{subfigure}[t]{\linewidth}
        \centering
        \includegraphics[width=.32\linewidth]{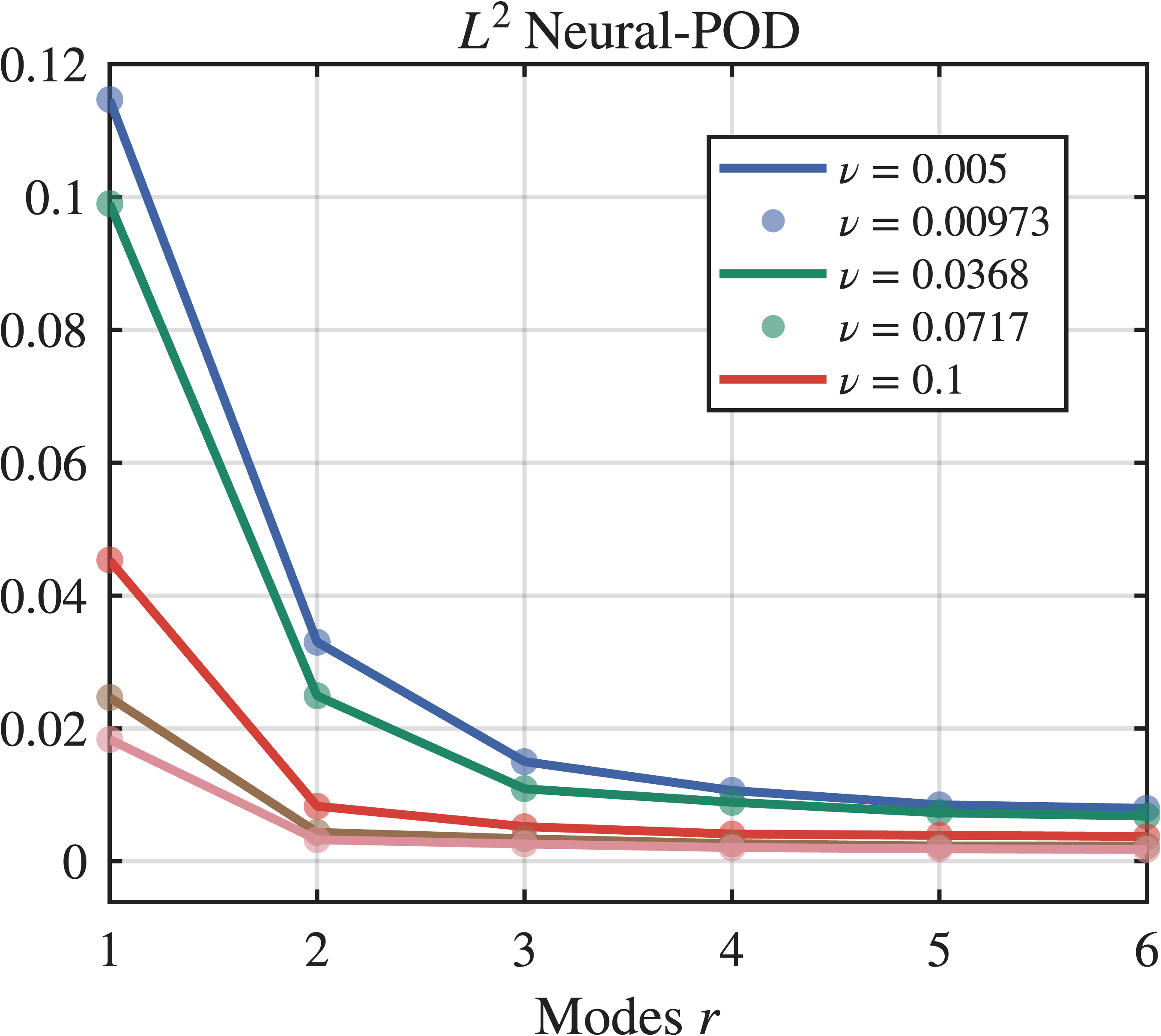}
        \hspace{1cm}
        \includegraphics[width=.32\linewidth]{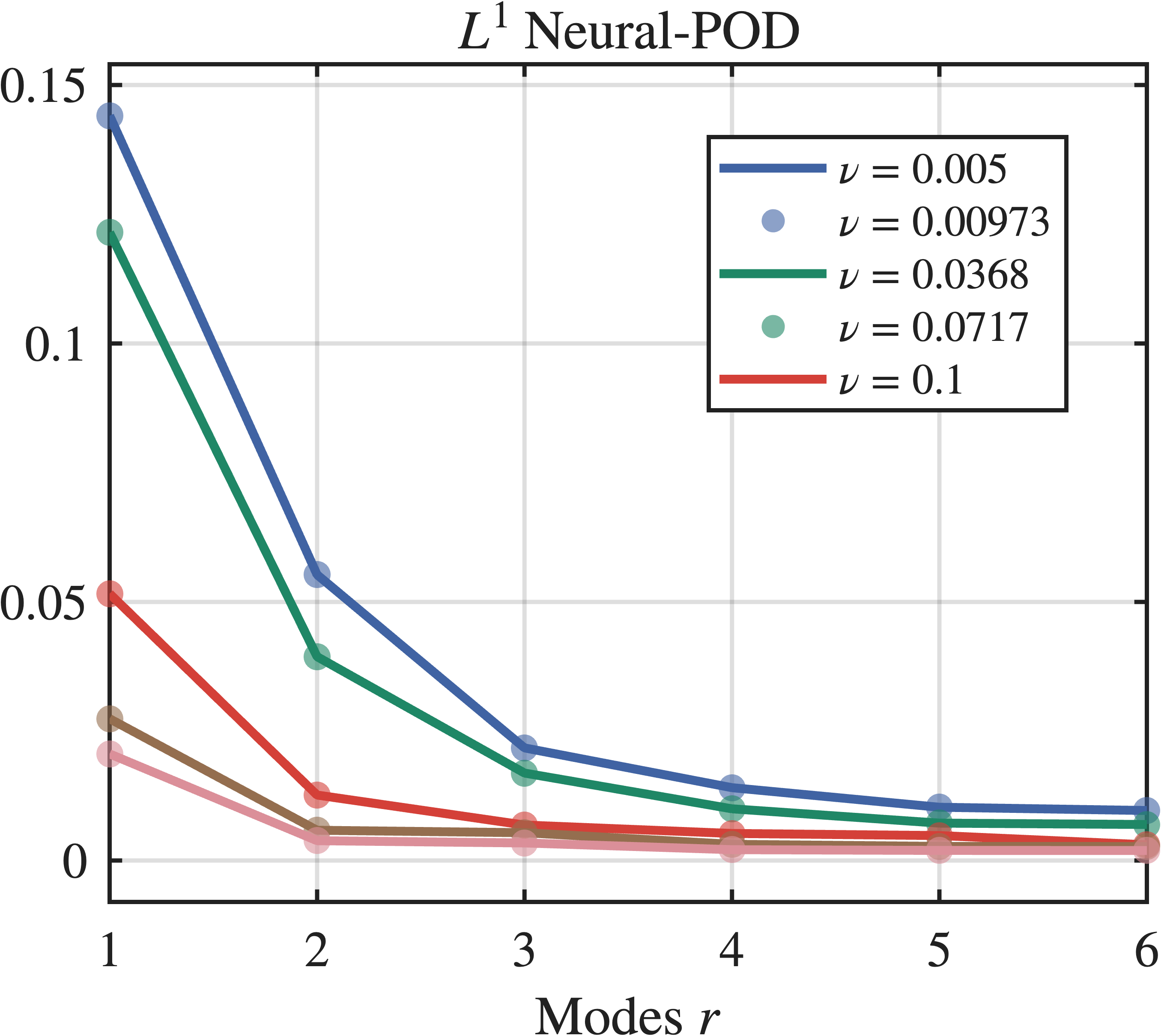}
        \caption{Neural-POD residuals versus modes $r$ for the $L^2$ loss (left) and the $L^1$ loss (right).}
        \label{fig:res-POD}
    \end{subfigure}

    \begin{subfigure}[t]{\linewidth}
        \centering

        \begin{subfigure}[b]{\linewidth}
          \centering
          \begin{subfigure}[b]{0.3\linewidth}
            \includegraphics[width=\linewidth]{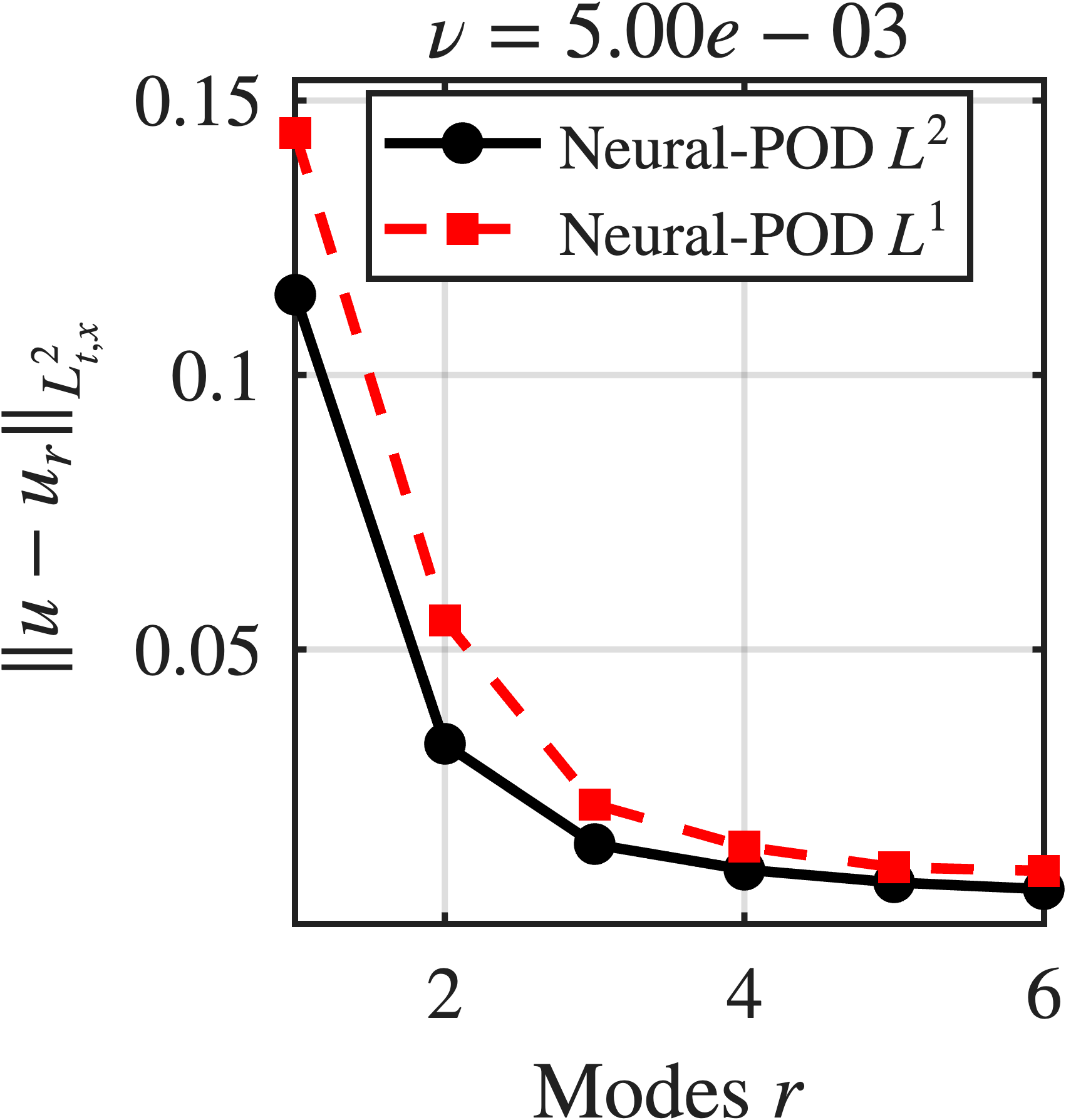}
          \end{subfigure}
          \begin{subfigure}[b]{0.3\linewidth}
            \includegraphics[width=\linewidth]{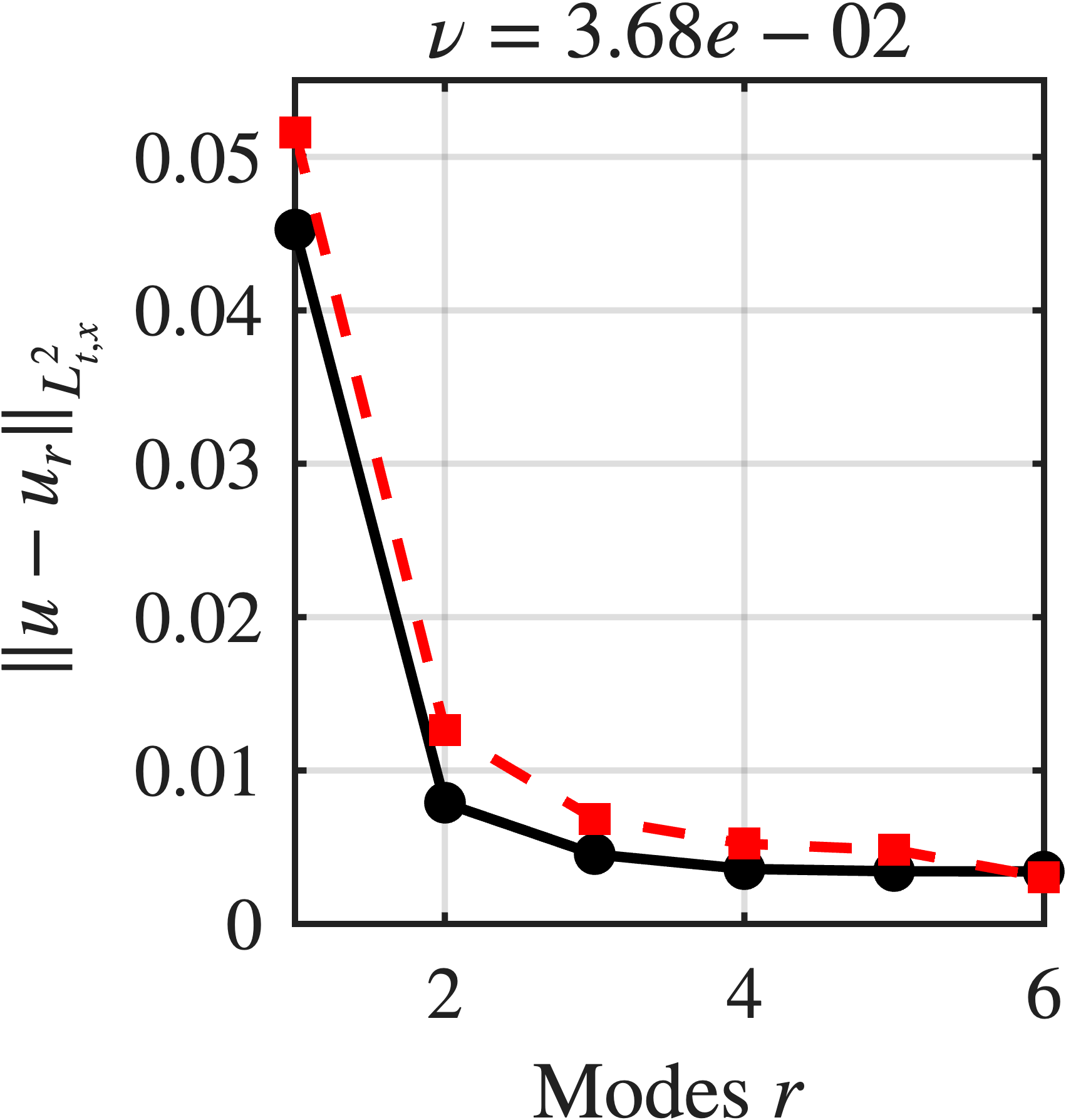}
          \end{subfigure}
          \begin{subfigure}[b]{0.3\linewidth}
            \includegraphics[width=\linewidth]{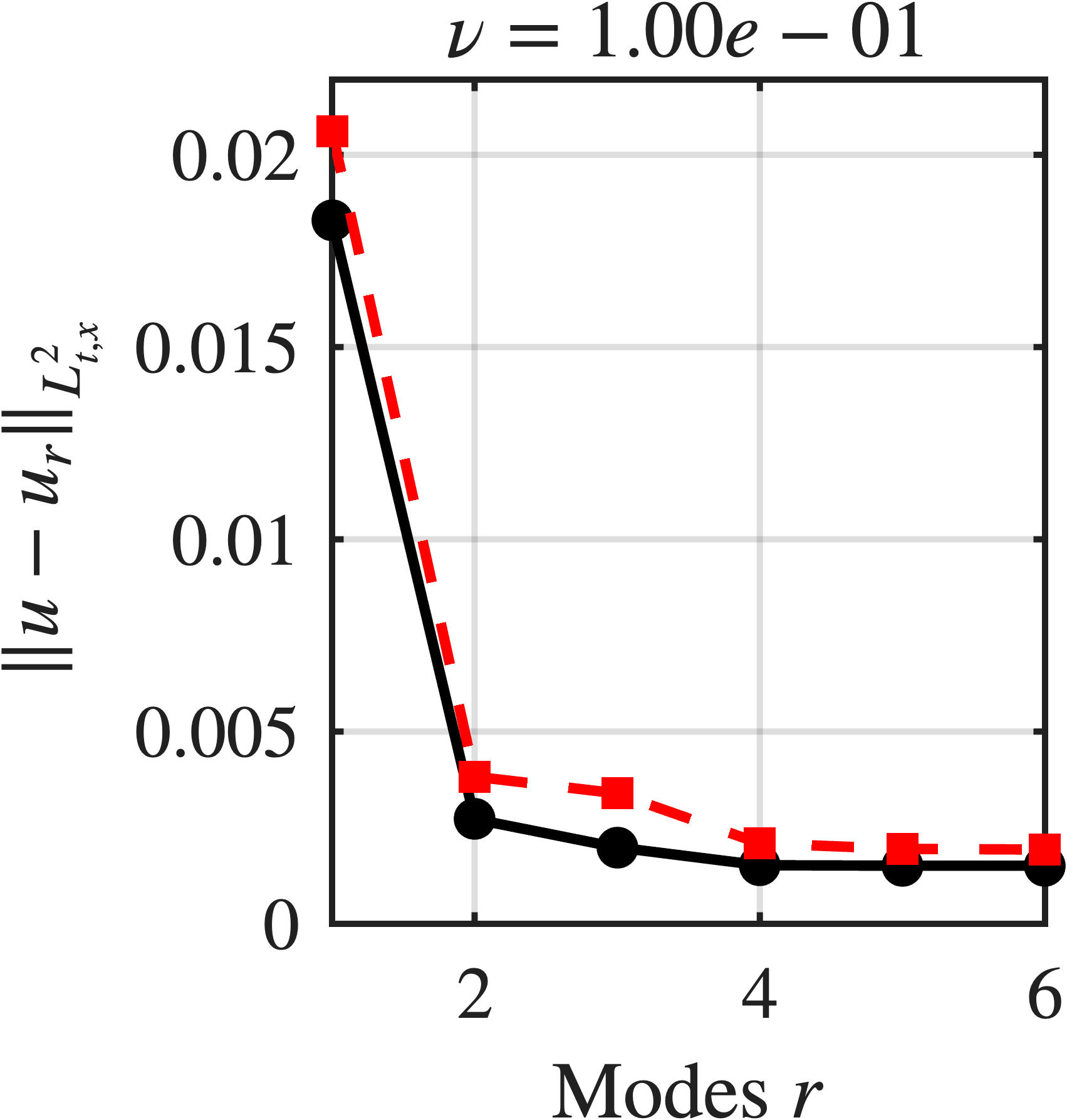}
          \end{subfigure}
          \caption{$L^{2}$ norm of the residuals for $L^{1}$- and $L^{2}$-optimal Neural-POD reconstructions.}
          \label{fig:pod-res-l1-12-a}
        \end{subfigure}

        \begin{subfigure}[b]{\linewidth}
          \centering
          \begin{subfigure}[b]{0.3\linewidth}
            \includegraphics[width=\linewidth]{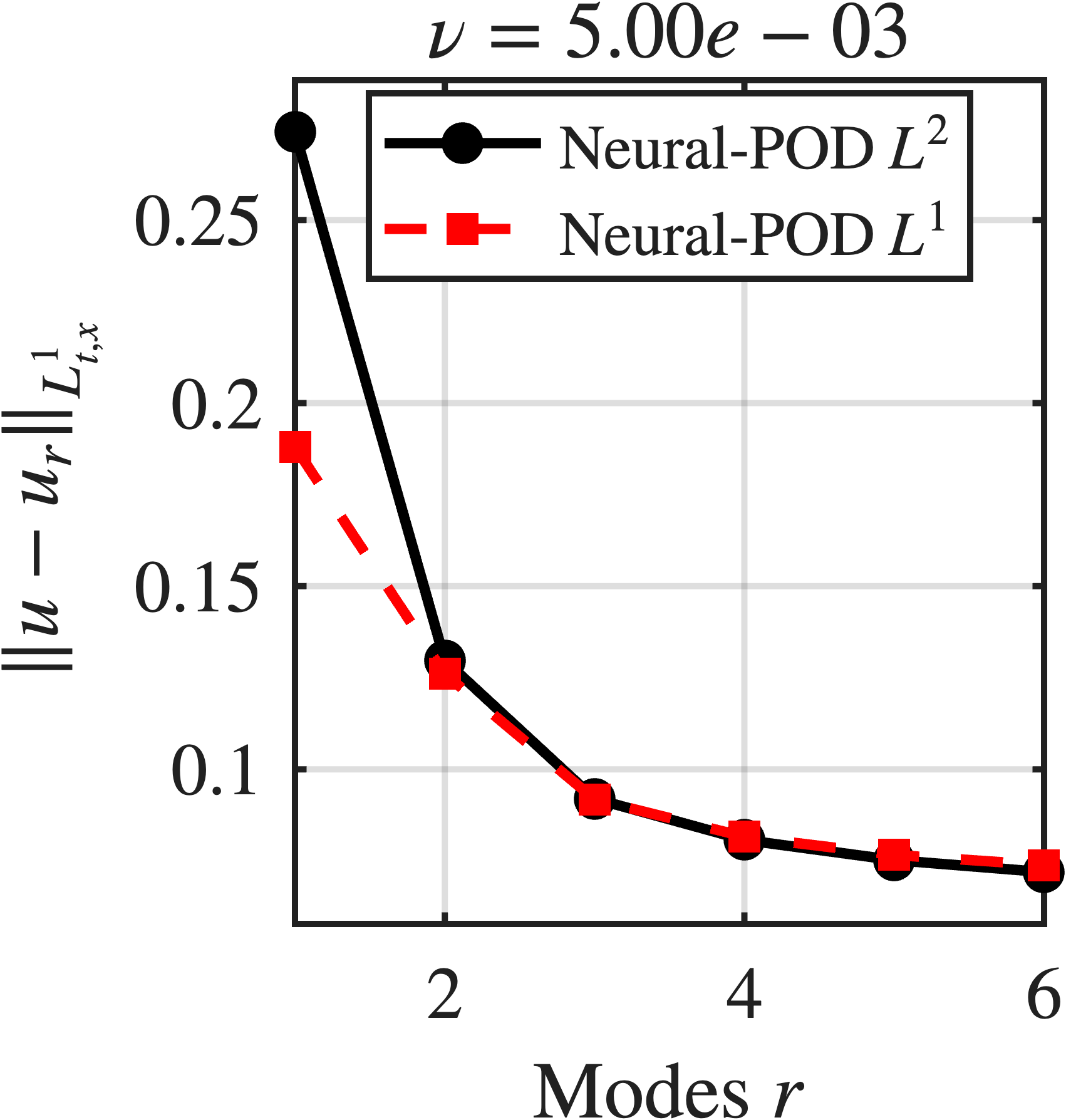}
          \end{subfigure}
          \begin{subfigure}[b]{0.3\linewidth}
            \includegraphics[width=\linewidth]{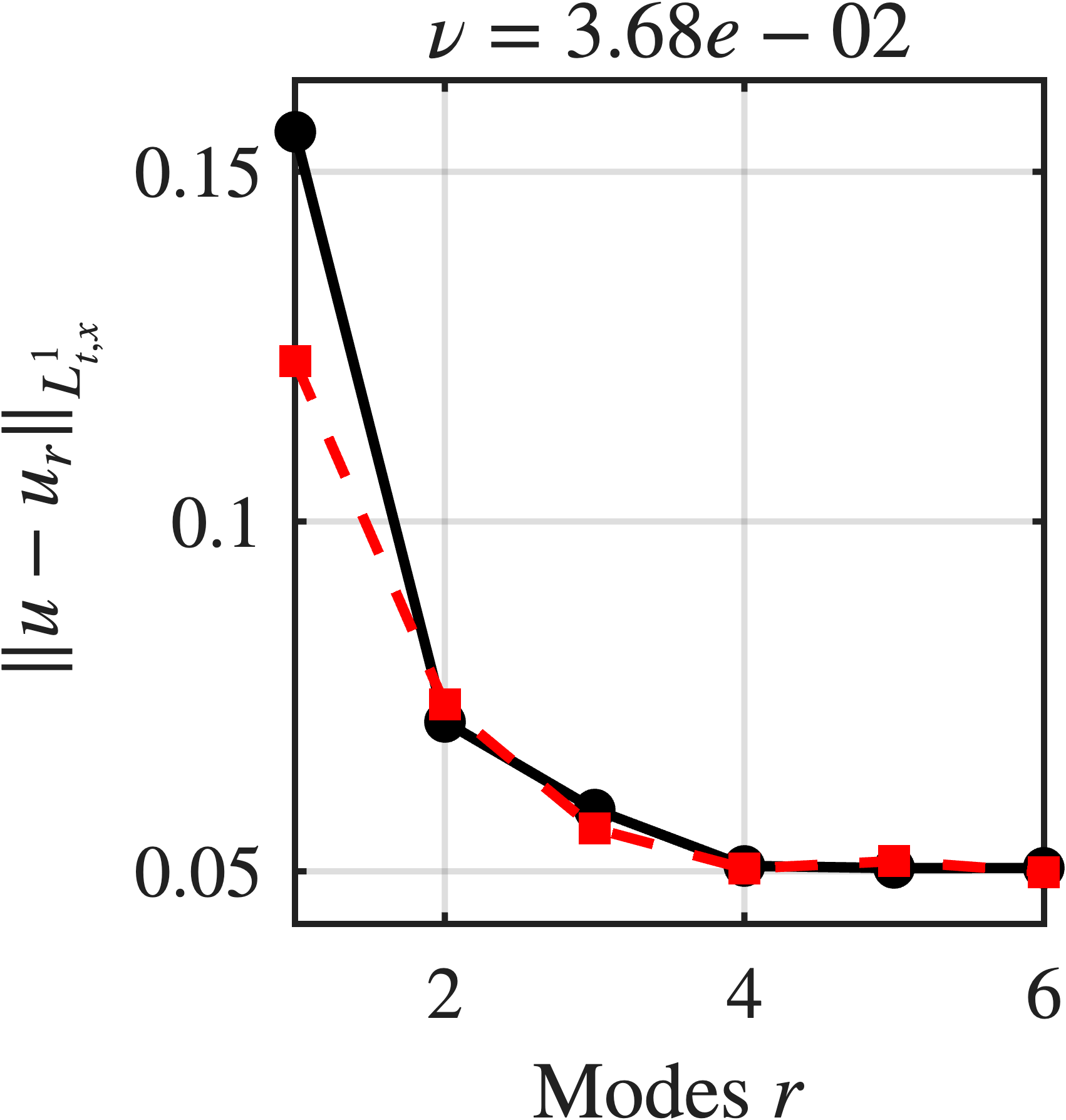}
          \end{subfigure}
          \begin{subfigure}[b]{0.3\linewidth}
            \includegraphics[width=\linewidth]{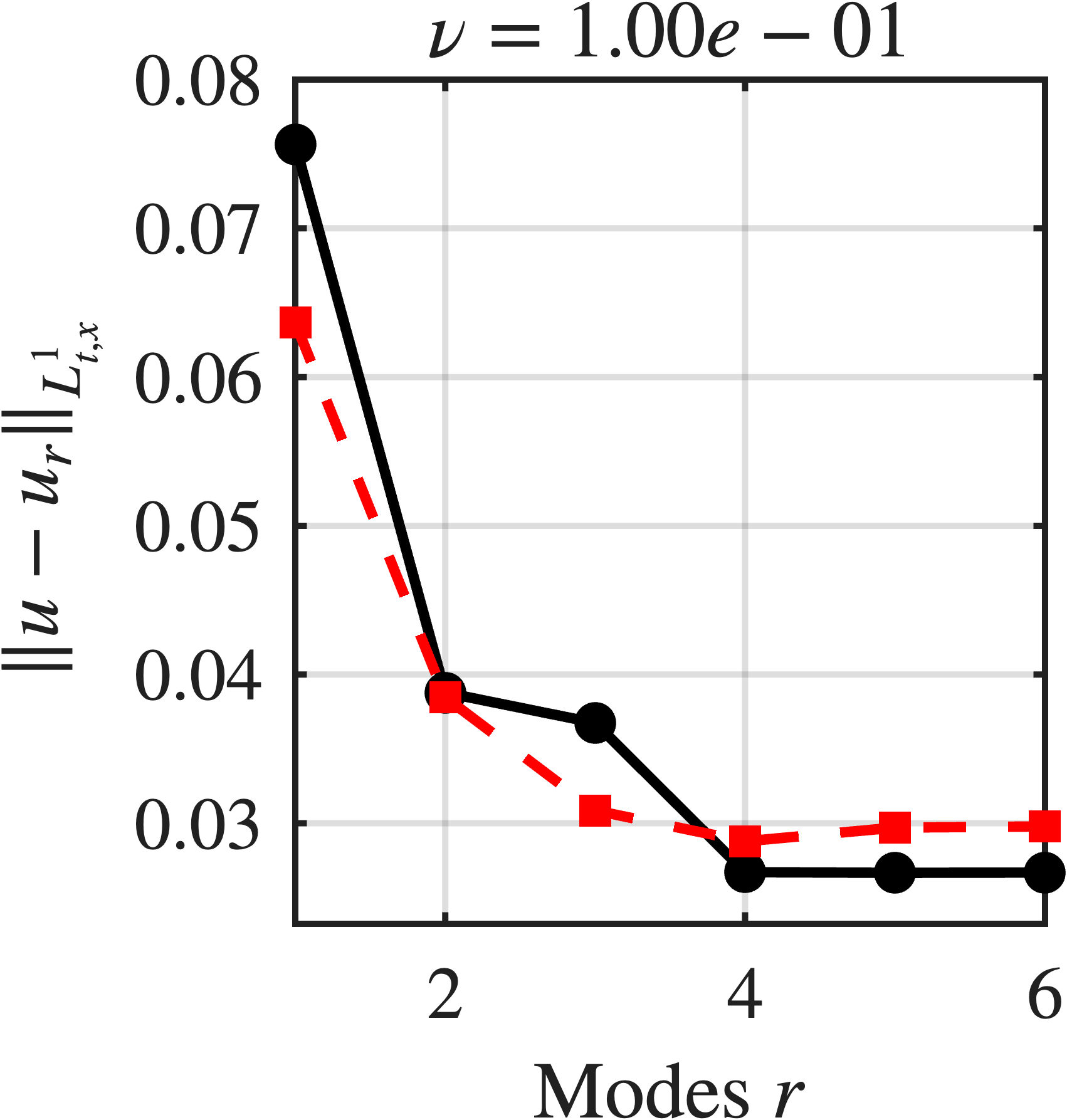}
          \end{subfigure}
          \caption{$L^{1}$ norm of the residuals for $L^{1}$- and $L^{2}$-optimal Neural-POD reconstructions.}
          \label{fig:pod-res-l1-12-b}
        \end{subfigure}

    \end{subfigure}
    \begin{subfigure}[t]{\linewidth}
        \centering
        \begin{subfigure}[b]{\linewidth}
            \centering
            \includegraphics[width=.8\linewidth]{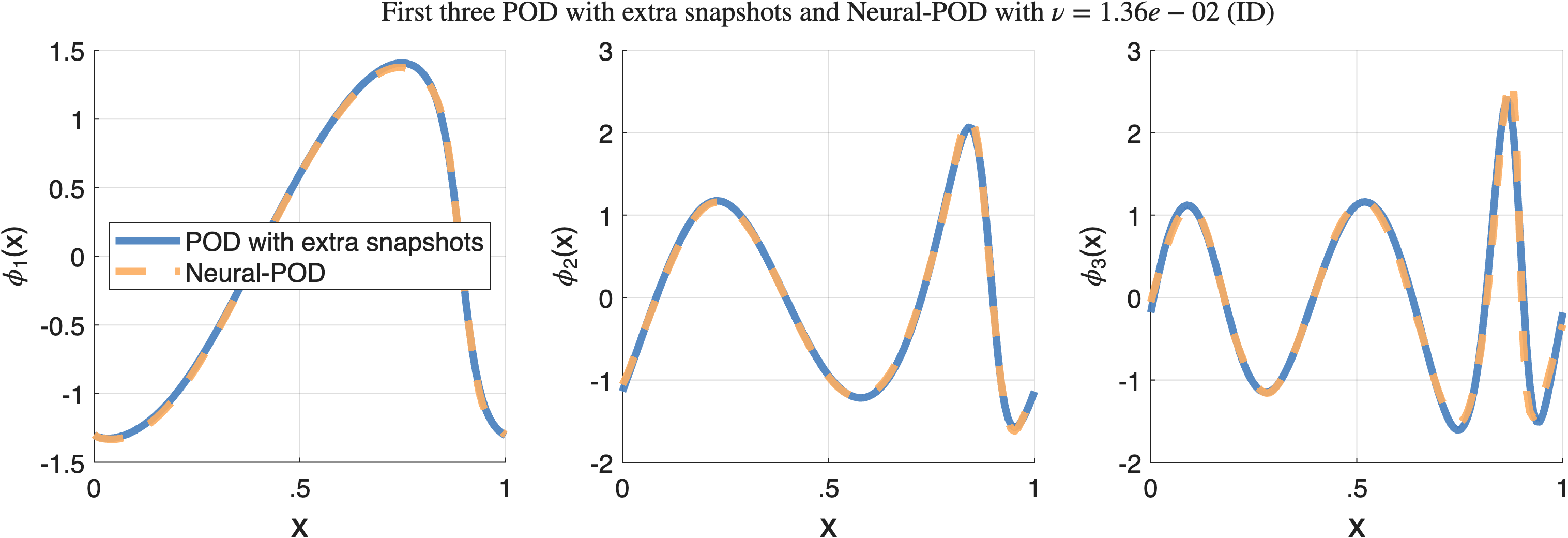}
            \caption{In-distribution results of the first three POD modes: comparison between traditional POD with extra snapshots and Neural-POD for an unseen viscosity value $\nu$. \label{fig:neural-pod-interp}}
        \end{subfigure}
        \begin{subfigure}[b]{\linewidth}
            \centering
            \includegraphics[width=.8\linewidth]{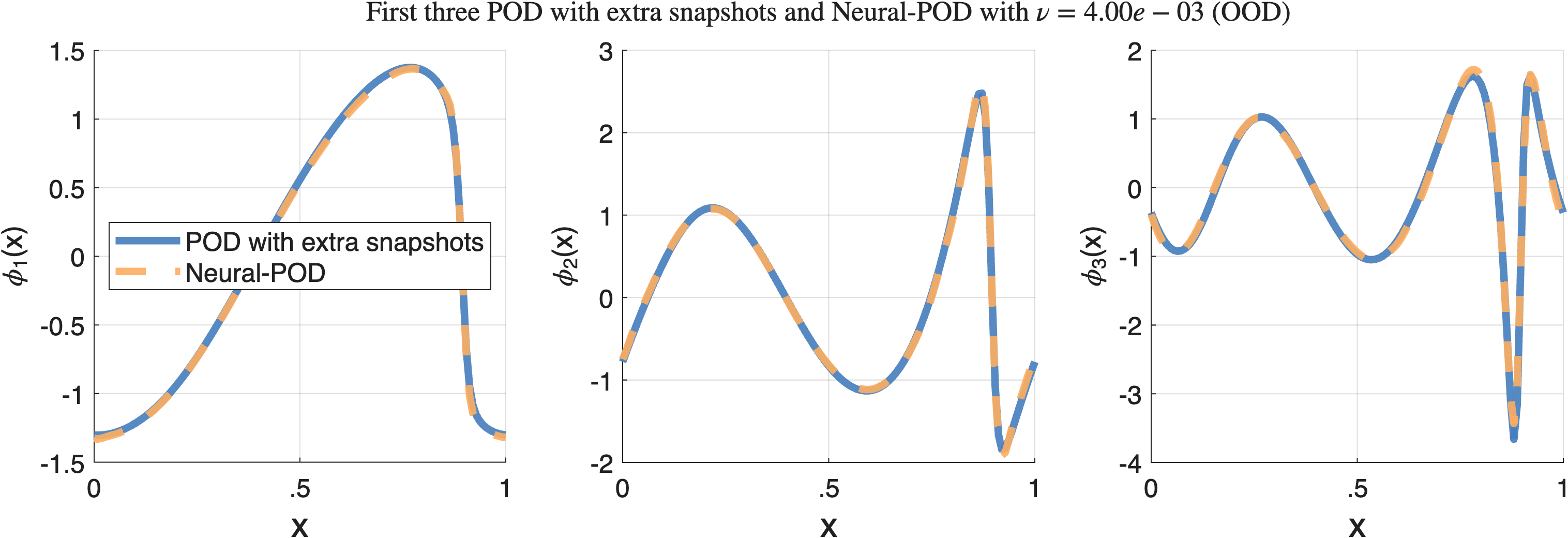}
            \caption{Out-of-distribution results of the first three POD modes: comparison between traditional POD with extra snapshots and Neural-POD for an unseen viscosity value $\nu$. \label{fig:neural-pod-extrap}}
        \end{subfigure}
        \label{fig:neural-POD-interp}
    \end{subfigure}
        \caption{Summary of Burgers Neural-POD results: (a) training residual decay versus mode index for $L^2$ and $L^1$ losses; (b--c) reconstruction residuals under $L^2$ and $L^1$ metrics across representative viscosities; (d--e) in-distribution and out-of-distribution comparisons of POD (with extra snapshots) and Neural-POD for an unseen viscosity $\nu$.}
\end{figure}

Extended Data Figure~\ref{fig:neural-POD-recon} illustrates the reconstruction of spatiotemporal fields using the Neural-POD model for a viscosity case not included in the training data.
The results are shown for latent dimensions of 2, 4, and 6.
Since the traditional POD basis depends on snapshot data from the training parameters, it cannot be constructed for this unseen case.
In contrast, the Neural-POD can accurately reconstruct the spatio-temporal dynamics even with only two or four modes which shows strong generalization capability and the ability to capture nonlinear correlations between spatial and temporal features beyond the training regime.
As the dimension of  modes increases, the Neural-POD reconstructions further improve and closely approach the truth solution, consistently yielding lower reconstruction errors and better preservation of localized fine-scale structures over time. 
For the case with $r = 6$, the model trained with the $L^1$ loss exhibits better performance compared to its $L^2$ loss Neural-POD, as it more effectively captures the steep gradients and localized features.
Figure \ref{fig:neural-POD} compares the first three spatial modes obtained from the classical POD and those learned through Neural-POD training using $L^1$- and $L^2$-based losses across different viscosity regimes.
The $L^2$-optimal Neural-POD tends to emphasize smooth, energy-dominant structures, while the $L^1$-based formulation better preserves localized sharp transitions by penalizing large pointwise deviations less severely.  Figure~\ref{fig:res-POD} shows that the Neural-POD residual decreases rapidly with the number of modes, closely mirroring the decay of the classical POD residual.

\FloatBarrier  
\clearpage
\thispagestyle{empty}
\clearpage

Figures~\ref{fig:pod-res-l1-12-a}--\ref{fig:pod-res-l1-12-b} compares the $L^1$ and $L^2$ norms of the reconstruction residuals obtained from Neural-POD models trained with $L^1$- and $L^2$-based loss functions across different viscosity values.
Panels~\ref{fig:pod-res-l1-12-a} and \ref{fig:pod-res-l1-12-b} correspond to the $L^2$ and $L^1$ residual norms, respectively, evaluated for three representative cases of $\nu=0.01$, $\nu=0.03$, and $\nu=0.05$.
In panel~\ref{fig:pod-res-l1-12-a}, the $L^2$-optimal Neural-POD produces smooth and globally accurate reconstructions, with residuals remaining low throughout the spatial domain, particularly for higher viscosity values where the flow is predominantly diffusive.
Panel~\ref{fig:pod-res-l1-12-b} presents the corresponding $L^1$ residual norms for both the $L^1$- and $L^2$-trained models.
The $L^1$-optimal Neural-POD exhibits improved accuracy in certain cases, particularly for specific reduced dimensions~$r$ ($r=1$ and $r=3$) where it better captures localized steep gradients and nonlinear features.
Because the $L^1$ norm penalizes large deviations less severely, it allows for sharper reconstructions in regions of strong nonlinearity, while the $L^2$ formulation maintains smoother but more globally averaged behavior.
Overall, the comparison indicates that the relative performance of the two loss formulations depends on both the viscosity parameter and the dimension~$r$.

\paragraph{Neural-POD for in-distribution and out-of-distribution  viscosity values}
In the following experiments, the Neural-POD model is trained under the same numerical setting as in Figure~\ref{fig:neural-pod-1}, using snapshot data generated at viscosity values $\nu$ sampled within the interval $[0.005,,0.01]$. This interval defines the training parameter domain for Neural-POD basis learning. 
Figures~\ref{fig:neural-pod-interp}--\ref{fig:neural-pod-extrap} show both the in-distribution and out-of-distribution performance of the Neural-POD framework.
Here, \textit{\textbf{in-distribution}} refers to evaluating the learned Neural-POD model at viscosity values $\nu$ within the parameter interval used for training, while \textit{\textbf{out-of-distribution}} refers to evaluating the model at $\nu$ values outside that interval.
More specifically, Figure \ref{fig:neural-pod-interp}  and  Figure \ref{fig:neural-pod-extrap}  compare the first three basis modes from traditional POD with those produced by Neural-POD trained using the $L^2$ loss. Because traditional POD constructs modes only at the discrete viscosity levels contained in the snapshot data, it cannot directly produce basis functions for intermediate or out-of-range parameter values. Neural-POD, on the other hand, learns a continuous mapping from $\nu$ to the modal space, allowing it to generate basis functions for both unseen in-distribution and out-of-distribution parameter regimes.
Notably, even for the fully unseen parameter value $\nu = 4\times10^{-3}$ (see Figure~\ref{fig:neural-pod-extrap}), which lies outside the training interval $[0.005,,0.01]$ and represents a $20\%$ \textit{\textbf{out-of-distribution}} shift below its lower bound, Neural-POD still generates modes that remain consistent with the corresponding POD modes that confirms its generalization capability.

\subsection*{Neural-POD for Navier-Stokes equations}
\begin{figure}[t]
    \centering

    \begin{subfigure}[t]{\linewidth}
        \centering
        \begin{subfigure}[t]{\linewidth}
            \centering
            \includegraphics[width=\linewidth]{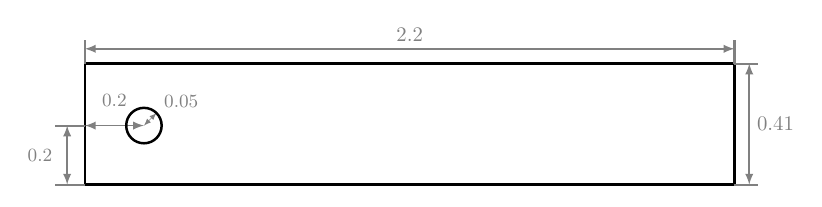}
            \caption{Geometry of two-dimensional flow past the cylinder.}
            \label{fig:cyldomain}
        \end{subfigure}

        \vspace{0.6em}

        \begin{subfigure}[t]{0.8\linewidth}
            \centering
            \includegraphics[width=\linewidth]{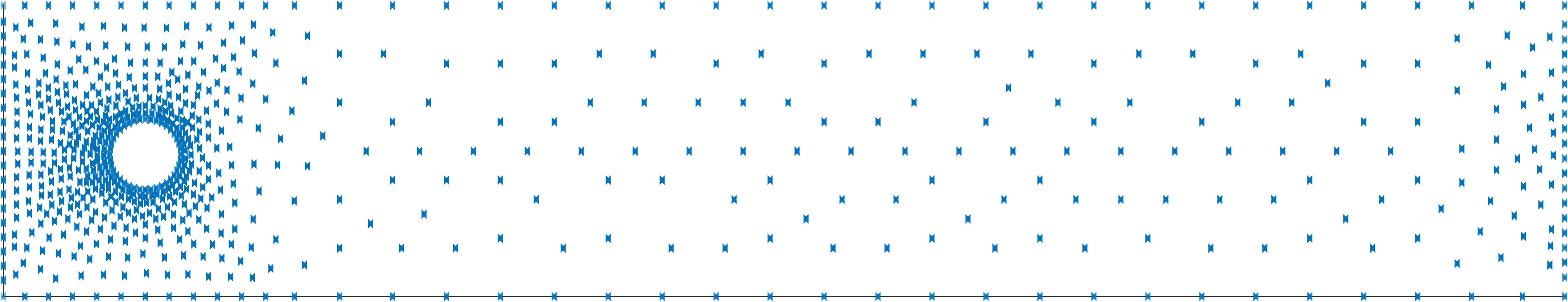}
            \caption{Sampling points for the flow past the cylinder test problem.}
            \label{fig:sample-cylinder}
        \end{subfigure}

    \end{subfigure}

    \begin{subfigure}[t]{\linewidth}
        \centering
        \begin{subfigure}[b]{\linewidth}
            \centering
            \includegraphics[width=\linewidth]{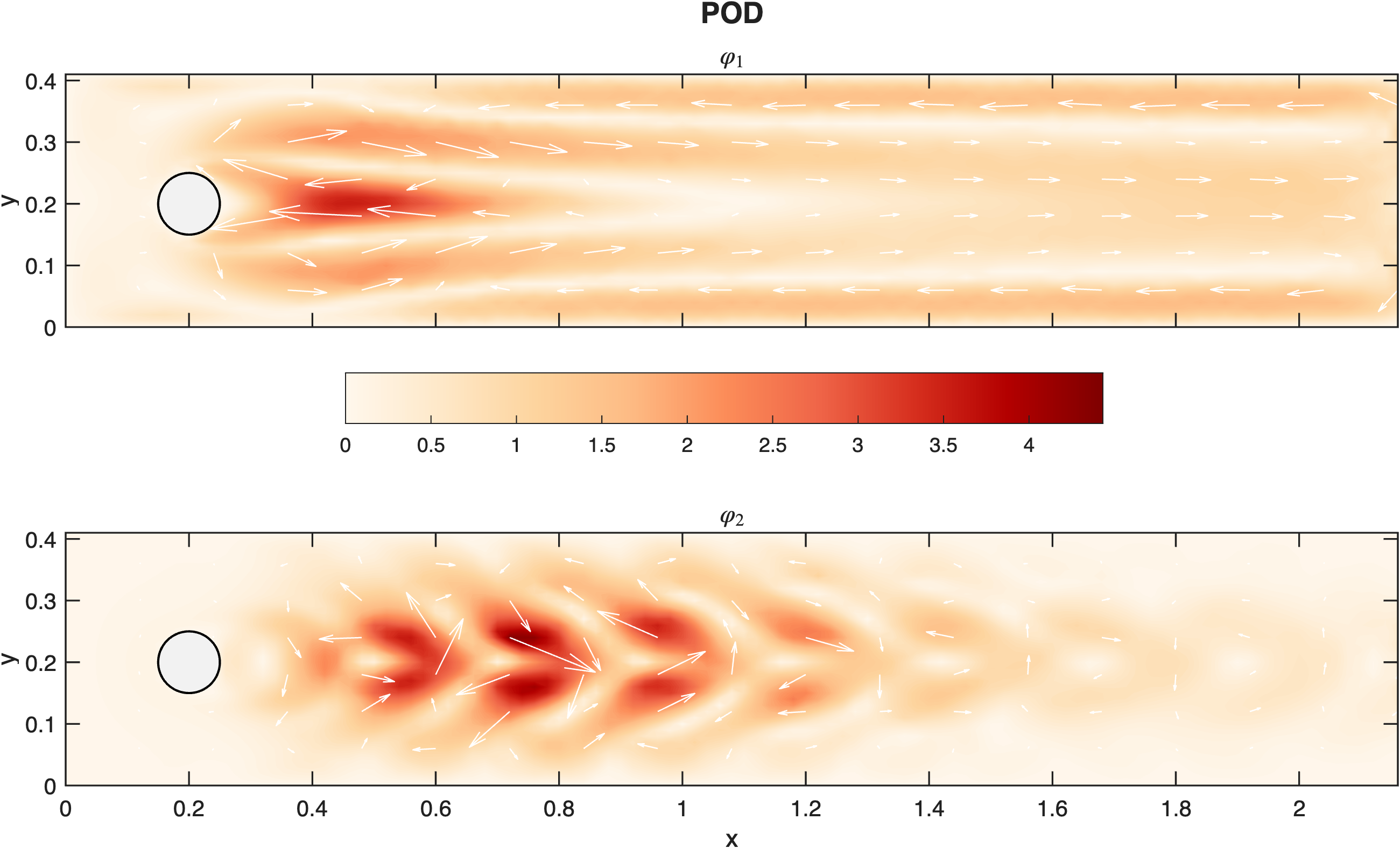}
            \caption{First two POD modes.}
            \label{fig:fcylinder-pod-a}
        \end{subfigure}

        \begin{subfigure}[b]{\linewidth}
            \centering
            \includegraphics[width=\linewidth]{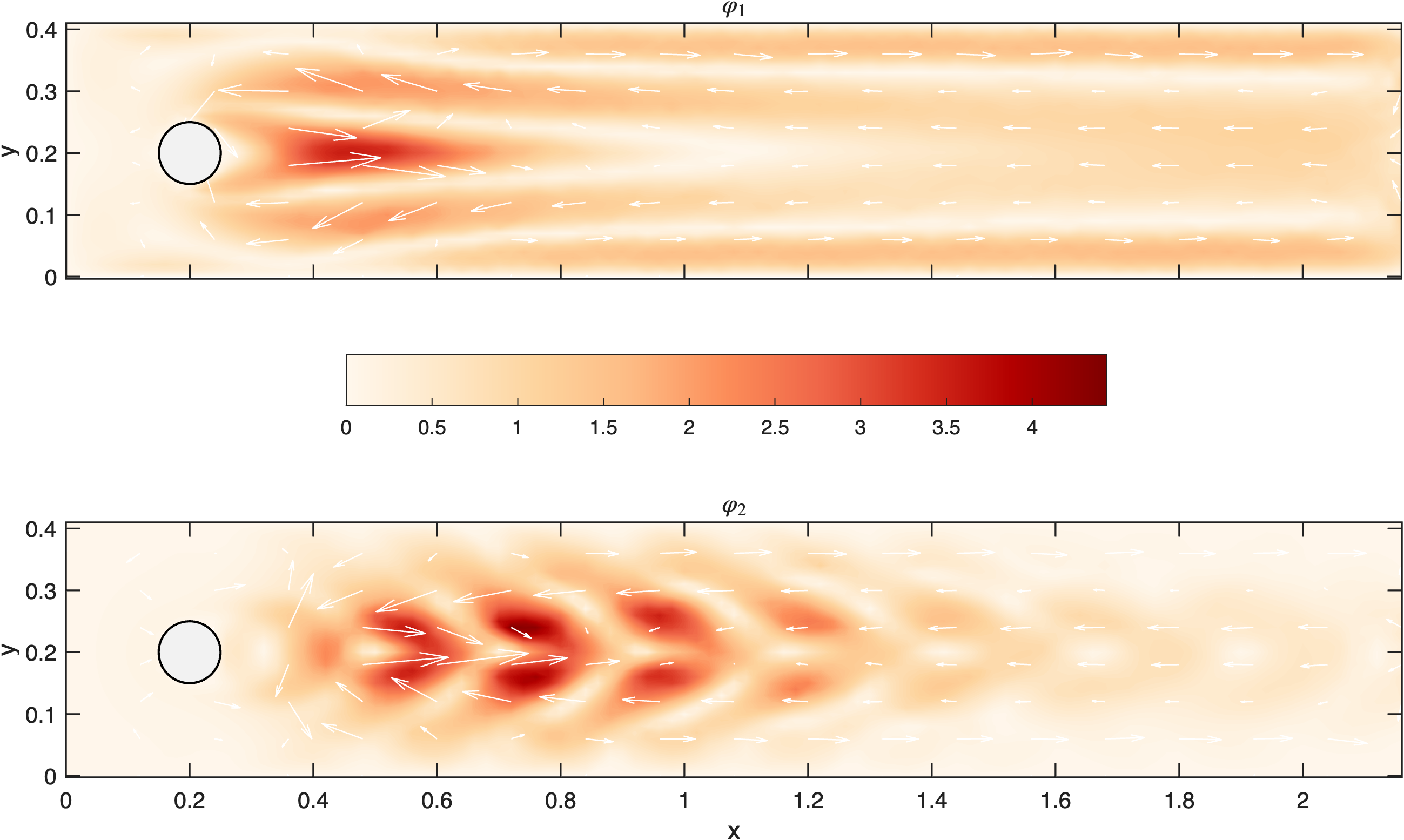}
            \caption{First two Neural-POD modes.}
            \label{fig:fcylinder-pod-b}
        \end{subfigure}
    \end{subfigure}

    \caption{(a--b) Geometry and sampling points of the two-dimensional flow past a cylinder problem. (c--d) First two modes of POD and Neural-POD for two-dimensional flow past the cylinder with $Re=50$.}
    \label{fig:flow-cylinder-and-modes}
\end{figure}

We also consider the Navier-Stokes equations (NSE) \cite{Lay02c,mou2021data,koc2022verifiability}:
\vspace*{-0.1cm}
 \begin{eqnarray}
     && \frac{\partial u}{\partial t}
     - Re^{-1} \Delta u
     + u \cdot \nabla u
     + \nabla p
     = {\bf 0} \, ,
     \label{eqn:nse-1} \\
     && \nabla \cdot u
     = 0 \, ,
     \label{eqn:nse-2}
	\\[-0.6cm]
	\nonumber
 \end{eqnarray}
 where $u$ is the velocity, $p$ the pressure, and $Re$ the Reynolds number. 
The computational domain is a $2.2\times 0.41$ rectangular channel with a radius $=0.05$ cylinder, centered at $(0.2,0.2)$, see Figures~\ref{fig:cyldomain}--\ref{fig:sample-cylinder}.  We prescribe no-slip boundary conditions on the walls and cylinder, and the following inflow and outflow profiles:
\begin{align}
u_{1}(0,y,t)&=u_{1}(2.2,y,t)=\frac{6}{0.41^{2}}y(0.41-y), \\ u_{2}(0,y,t)&=u_{2}(2.2,y,t)=0,
\end{align} 
where $u=\langle u_1, u_2 \rangle$.

The dataset used is generated by a high-fidelity finite element simulation of the unsteady Navier–Stokes equations at $Re=50$. The spatial discretization employs a Taylor–Hood element pair (quadratic velocity, linear pressure) on an unstructured triangular mesh that provides sufficient resolution of the boundary layers and wake structures. The resulting finite element discretization yields approximately $1.2\times10^4$ velocity degrees of freedom (DOFs).
Temporal integration is performed using a second-order backward differentiation formula (BDF2). 
A total of $1000$ velocity snapshots are recorded uniformly in time over the interval $[10,20]$. To make a consistent comparison, these snapshots are used to generate both the Proper Orthogonal Decomposition (POD) basis and to train the Neural–POD.
Figures~\ref{fig:fcylinder-pod-a}--\ref{fig:fcylinder-pod-b} illustrates the first two spatial modes identified by the classical POD and Neural–POD. The close agreement between the two confirms the consistency of Neural–POD with the traditional POD framework.

\subsection*{Neural-POD in Reduced Order Modeling}
In this section, we investigate the Neural-POD applied in Galerkin projection reduced order models (Galerkin ROM) in the numerical simulation of the one-dimensional 
viscous Burgers equation in equations \eqref{eq:burgers} with the initial condition:
\begin{align}
\begin{aligned}
    & u(x,0) = u_0(x) = 
\left\{
\begin{aligned}
    & 1, && x \in (0, 1/2], \\
    & 0, && x \in (1/2, 1],
\end{aligned}
\right.
\end{aligned}
\label{eq:burgers:ic}
\end{align}
and $\nu = 10^{-3}$. This test problem has been used in \cite{mou2021data,koc2023uniform}.

\paragraph{Snapshot Generation} The truth data are obtained using a linear finite element spatial discretization with mesh size $h = 1/512$ and a Crank--Nicolson time discretization with timestep size $\Delta t = 10^{-4}$. A total of $101$ snapshots are collected at equally spaced time instances over the interval $t \in [0, 1]$ for constructing the POD and Neural-POD modes. More details are provided in Supplementary Section A.2. and Supplementary Data, Fig. A3.
\clearpage
\thispagestyle{empty}
\clearpage


\paragraph{Discussion of Numerical Results} 
Figure \ref{fig:pod-neural-burgers-comparison} compares the modes from classical POD and Neural–POD under different optimality criteria for Burgers equation with initial condition in \eqref{eq:burgers} . The $L^2$-optimal Neural–POD modes closely resemble the classical POD modes, showing that the neural representation preserves the same energy-optimization structure when trained in the same norm. In contrast, the $L^1$-optimal Neural–POD produces sharper and more localized modes, emphasizing regions with strong gradients or discontinuities. This indicates that while the $L^2$-trained Neural–POD recovers the traditional POD basis, the $L^1$ formulation provides a flexible alternative  for capturing nonlinear and nonsmooth dynamics.
We consider two projection-based ROMs: the standard POD-ROM, which uses classical POD modes, and the Neural-POD-ROM, which employs modes learned via Neural-POD. 

\begin{table}[!htp]
\caption{
$L^2$ errors of POD--ROM and Neural--POD--ROM for the one-dimensional Burgers equation.
At $\nu=0.002$, training snapshots are available for both models.
At $\nu=0.005$, Neural--POD--ROM is evaluated in a prediction setting without training snapshots.
}
\label{tab:podrom-combined}
\begin{tabular*}{\textwidth}{@{\extracolsep\fill}clccc}
\toprule
$\nu$ & Model & $r=2$ & $r=4$ & $r=6$ \\
\midrule
\multirow{2}{*}{0.002}
 & POD--ROM                 & 2.10e-01 & 1.66e-01 & 1.76e-01 \\
 & Neural--POD--ROM ($L^2$) & 2.11e-01 & 1.67e-01 & 1.77e-01 \\

\midrule
\multirow{2}{*}{0.0025}
 & POD--ROM                 & \multicolumn{3}{c}{\textit{\textbf{No in distribution ability}}} \\
 & Neural--POD--ROM ($L^2$) & \textbf{9.61e-02} & \textbf{1.18e-01} &  \textbf{1.35e-01}\\
\midrule
\multirow{2}{*}{0.005}
 & POD--ROM                 & \multicolumn{3}{c}{\textit{\textbf{No out of distribution ability}}} \\
 & Neural--POD--ROM ($L^2$) & \textbf{1.91e-01} & \textbf{1.21e-01} & \textbf{8.46e-02} \\
\botrule
\end{tabular*}
\end{table}
Table \ref{tab:podrom-combined} compares the $L^2$ reconstruction errors of the standard POD–ROM and the proposed Neural–POD–ROM for the one-dimensional Burgers equation for viscosity values $\nu=0.002$ with different ROM dimensions $r$. 
At this parameter value, where training snapshots are available, both POD-ROM and Neural-POD-ROM achieve comparable accuracy for all different ROM dimensions. This confirms that, \emph{when training data are available at the given parameter value}, the neural representation of basis functions matches the ability of classical POD to capture the dominant coherent structures.
Table~\ref{tab:podrom-combined} presents ROM results at the unseen viscosity value $\nu = 0.005$, which is not included in the training snapshots. In this regime, the classical POD–ROM cannot be constructed due to \emph{the absence of simulation snapshots} at the target parameter value. In contrast, the Neural–POD–ROM can still be formulated using pretrained Neural–POD predictions even when the PDE parameter, i.e., the viscosity $\nu$, is not included in the training set, and it continues to produce accurate reconstructions for all ROM dimensions.

\subsection*{Neural-POD in Deep Operator Learning}



We consider the one-dimensional Burgers’ equations~\eqref{eq:burgers-a}--\eqref{eq:burgers-c} following the same configuration described in Section~\ref{sec:neural-pod-numeric}.
The dataset is adopted from~\cite{lu2022comprehensive}, where the initial conditions are sampled from a Gaussian random field characterized by a Riesz kernel,
$\mu = \mathcal{R}(0,,625(-\Delta + 25I)^{-2})$,
with $\mu$ denoting the probability measure over the function space and $\Delta$ and $I$ representing the Laplacian and identity operators, respectively.
Both input and output fields are discretized on a uniform grid with 128 spatial points.

In contrast to the method proposed in~\cite{lu2022comprehensive}, which employs snapshots at $t=1$ to construct the POD basis, our Neural-POD framework learns the basis functions directly from the temporal evolution of the Burgers’ equation across multiple time steps.
We employ four latent modes in the reduced representation and train the model on 128-point spatial grids, while evaluating its reconstruction and generalization performance on coarser 64-point grids.
This setup enables assessment of both the model’s spatial resolution robustness and its capacity to generalize across varying discretizations and viscosity parameters. 
This is further confirmed by Extended Data Tables~\ref{tab:compare_pod_models-2}.
At the coarse resolution ($N=64$), the Neural-POD–DeepONet achieves lower testing errors than the traditional POD–DeepONet, with roughly a $20\%$ reduction in both $L_1$ and $L_2$ metrics.
These results, consistent with Extended Data Figure~\ref{fig:npod-deeponet}, show that the Neural-POD framework offers clear benefits under limited resolution, while recovering classical POD behavior as the discretization becomes sufficiently fine.
At coarse resolution ($N=32$), the Neural-POD–DeepONet exhibits noticeably improved reconstructions over the traditional POD–DeepONet, capturing sharper transitions.
This improvement can be attributed to the adaptive and data-driven nature of the Neural-POD basis, which enhances the representation of localized features under limited spatial resolution.
As the training resolution increases to $N=64$, the difference between the two approaches becomes less pronounced, though the Neural-POD still achieves slightly lower reconstruction errors.
For the finest resolution ($N=128$), both Neural-POD–DeepONet and POD–DeepONet produce nearly similar results, indicating that the advantage of the learned basis diminishes as the resolution becomes sufficiently high to resolve the dominant flow structures.
In this regime, the $L^1$-based Neural-POD performs marginally worse than the $L^2$ variant due to its less stringent global smoothness constraint.
Overall, the results suggest that the Neural-POD framework provides clear advantages at coarse or moderately resolved training settings, where it compensates for the loss of spatial detail through adaptive modal learning, while converging to the classical POD behavior at higher resolutions. 
Extended Data Figure~\ref{fig:npod-deeponet-error} shows the comparison between the proposed Neural-POD–DeepONet and the traditional POD–DeepONet with different spatial training resolutions, $N=32$, $N=64$, and $N=128$.
Each model is evaluated on unseen initial conditions of the Burgers’ equation to verify its ability to generalize for both data realizations and discretization levels.

\section*{Discusssions \label{sec:conclusion}}
In this work, we introduced the \textit{Neural Proper Orthogonal Decomposition} (Neural-POD), a neural operator approach that generalizes the classical POD through nonlinear basis learning with neural networks.
Unlike traditional linear POD, which relies on a fixed linear subspace derived from singular value decomposition, Neural-POD constructs orthogonal basis functions via iterative residual minimization using trainable neural representations.
This neural formulation enables basis optimization under arbitrary norms, such as $L^1$, $L^2$, and naturally accommodates nonlinear and multiscale dynamics captures nonlinear spatiotemporal features.
Numerical tests are performed on benchmark problems, including the one-dimensional Burgers’ equation and the two-dimensional Navier–Stokes equations, which shows the robustness of the proposed framework.
We also show that Neural-POD can be effectively used in projection-based reduced order models (ROMs), and deep operator network (DeepONet) with numerical tests.   
The pretrained and plug-and-play properties of Neural-POD also creates a promising pathway toward an open-source, community-driven database, in which shared Neural-POD basis libraries may further facilitate reproducibility and further expand the applicability in communities of ROM and operator learning.

Future research will focus on the following different directions: First, we will extend the Neural-POD framework to more challenging and high-dimensional scientific problems such the turbulent three-dimensional Navier–Stokes or Boussinesq equations, where nonlinear modal interactions and multi-scale energy cascades provide a demanding test \cite{mou2023energy,bright2013compressive,lin2009efficient}.
Another promising direction is the integration of Neural-POD with physics-informed neural networks (PINNs) to enforce physical constraints during basis construction, thereby improving interpretability and extrapolation for physics-guided model reduction \cite{cai2021physics,lu2025mopinnenkf}.
Furthermore, coupling Neural-POD with data assimilation techniques such as Ensemble Kalman Filters (EnKF) or variational methods may enable real-time state estimation and model correction in dynamical systems \cite{popov2021multifidelity,bilionis2013multi}.
Also, in digital-twin applications, Neural-POD can be employed to build efficient and adaptive surrogate models that enable real-time monitoring, prediction, and control of complex physical systems \cite{xiu2025computational,jones2020characterising}. Lastly, we plan to extend the framework to parameterize and learn geometrical variations that enables the community to contribute new mesh configurations and expand the applicability of Neural-POD to complex domains with heterogeneous geometries \cite{shukla2024deep}.

\section*{Methods \label{sec:neural-pod} }

\subsection*{Classic Proper Orthogonal Decomposition (POD)}
The \textit{classical} Proper Orthogonal Decomposition (POD) is a widely used technique for extracting low-dimensional structures from high-dimensional spatiotemporal data \cite{berkooz1993proper,chatterjee2000introduction}. It is especially prominent in fluid dynamics, model reduction, and system identification due to its mathematical rigor, computational efficiency, and optimality properties in the $L^2$ norm.

Given a set of $M$ snapshots ${u(x,t_j)}{j=1}^M$ collected from a high-dimensional system, POD seeks a set of orthonormal basis functions ${\phi_i(x)}{i=1}^r$ that best approximate the snapshots in a least-squares sense. Formally, POD solves the following minimization problem:
\begin{equation}
\min_{\{\phi_i\}_{i=1}^r} \frac{1}{M} \sum_{j=1}^M \left\| u(x, t_j) - \sum_{i=1}^r \langle u(\cdot, t_j), \phi_i \rangle \phi_i(x) \right\|_{L^2}^2,
\end{equation}
subject to the orthonormality constraint $( \phi_i, \phi_j ) = \delta_{ij}$. The solution is obtained via singular value decomposition (SVD) of the snapshot matrix, and the resulting modes ${\phi_i}$ span an $r$-dimensional subspace that captures the most energetic components of the dataset \cite{gubisch2017proper,kunisch2008proper,berkooz1992observations}.
POD guarantees optimality in the $L^2$ sense: for any given rank $r$, the POD basis minimizes the projection error among all $r$-dimensional linear subspaces. This makes it a powerful tool for linear dimensionality reduction and the foundation for many reduced order models (ROMs), where the governing equations are projected onto the POD subspace using Galerkin or Petrov–Galerkin projection \cite{girfoglio2021pod,iliescu2013variational,mou2020data,tsai2025time}.

Despite its strengths, classical POD has several limitations. Because its basis is strictly linear, it often struggles to represent complex nonlinear features, sharp gradients, or discontinuities \cite{volkwein2013proper,iliescu2014are}. Furthermore, the POD basis is highly dependent on the resolution and parameter regime of the snapshot data, limiting its generalizability \cite{mifsud2016variable,roderick2014proper}. These limitations (see left panel of Figure \ref{fig:neural_vs_pod}) motivate the development of more flexible, data-driven alternatives, such as the \textbf{\textit Neural-POD} framework introduced in this work (see right panel of Figure \ref{fig:neural_vs_pod}). 
\subsection*{Neural-POD Framework}
The Neural Proper Orthogonal Decomposition (Neural-POD) in this work provides a data-driven framework for constructing low-dimensional representations of high-dimensional spatiotemporal data. Unlike traditional POD methods that employ linear decompositions via singular value decomposition (SVD), Neural-POD utilizes neural networks to learn flexible spatial and temporal basis functions that are capable of capturing nonlinear structures in the data. Algorithm \ref{alg:neural-pod} provides a sketch of the Neural-POD.

The algorithm starts with the input of a dataset $\mathcal{D}_u = \{\mathcal{D}_u^{(i)}\}_{i=1}^N$, where each snapshot $\mathcal{D}_u^{(i)}$ consists of a time point $t^{(i)}$ and a collection of $M$ spatial-temporal samples $\{(x^{(i,j)}, u^{(i,j)})\}_{j=1}^M$. A small tolerance parameter $\text{Tol} > 0$ is also provided to determine the desired approximation accuracy. The objective is to iteratively decompose the data into a sum of separable spatiotemporal modes parameterized by neural networks until the residual error falls below the specified tolerance.
The process begins by initializing the index $p = 0$ and training a neural network $\Phi_0(x; \theta_0)$ that minimizes the mean $L^2$ error between neural network output and the provided data $u(x,t)$, sampled over the entire dataset. This first Neural-POD serves as the initial approximation to the solution, often capturing the dominant spatial pattern. The residual $r(x,t)$ is then computed as the difference between the original data and the initial approximation.
If the residual error, measured in the $L^2$ sense over all $(x,t)$ pairs, is within the tolerance, the algorithm proceeds to iteratively add new modes. In each iteration, $p$ is incrementally increased, i.e., $p\to p+1$, and two neural networks $\Phi_p(x; \theta_p)$ and $\Psi_p(t; \beta_p)$ are randomly initialized. These networks are trained jointly to approximate the residual $r(x,t)$ by the product $\Phi_p(x; \theta_p)\Psi_p(t; \beta_p)$. After training, the residual is updated by subtracting the new mode from it.
Upon completion, the algorithm returns the number of modes $P$, the initial approximation $\Phi_0(x; \theta_0)$, and the set of learned spatial and temporal basis functions $\{\Phi_p(x; \theta_p), \Psi_p(t; \beta_p)\}_{p=1}^P$. This decomposition provides a compact, interpretable, and expressive representation of the original system, which can be further used for reduced order modeling. 

Beyond being a standalone data-driven method, Neural-POD can also be interpreted as a bridge between classical projection-based reduced order modeling (ROM) and modern operator learning frameworks. As illustrated in Figure~\ref{fig:bridge_neural_pod}, the learned neural representations of POD basis functions can be directly used in Galerkin projection-based ROMs, while simultaneously functioning as pretrained, interpretable branch networks within DeepONet architectures.



\begin{algorithm}[h]
\caption{Neural Proper Orthogonal Decomposition (Neural-POD)\label{alg:neural-pod}}
\begin{algorithmic}[1]
  \State \textbf{Input:}
  \Statex \quad $\displaystyle \mathcal{D}_u \;=\;\Bigl\{\,\mathcal{D}_u^{(i)}\Bigr\}_{i=1}^N,\quad
     \mathcal{D}_u^{(i)} = \Bigl(\,t^{(i)},\,\bigl\{\,x^{(i,j)},\,u^{(i,j)}\,\bigr\}_{j=1}^M\Bigr)$ 
    \quad // $N$ snapshots of $u(x,t)$
  \Statex \quad $0 < \text{Tol} \ll 1$ 
    \quad // small tolerance for target accuracy

  \State \textbf{Output:}
  \Statex \quad $P,\;\Phi_{0}(\,\cdot\,;\theta_0),\;\{\,\Phi_{p}(\,\cdot\,;\theta_p),\,\Psi_{p}(\,\cdot\,;\beta_p)\}_{p=1}^P$
    \quad // a set of $2P+1$ parameterized basis functions

  \State $p \gets 0$
  
  \State $\displaystyle \theta_0 \;\gets\;
    \arg\min_{\theta_0}\;\sum_{(x,t)\sim\mathcal{D}_u}\,
      \bigl\|\,u(x,t)\;-\;\Phi_{0}\bigl(x;\theta_0\bigr)\bigr\|_{2}^2$
  
  \State $r(x,t) \;\gets\; u(x,t)\;-\;\Phi_{0}\bigl(x;\theta_0\bigr)$

  \While{$\displaystyle \sum_{x,t}\bigl[r^2(x,t)\bigr]\;\ge\;\text{Tol}$}
    \State $p \gets p+1$
    \State \textbf{Randomly initialize} new neural networks 
      $\Phi_p\bigl(x;\theta_p\bigr),\;\Psi_p\bigl(t;\beta_p\bigr)$
      
    \State $\displaystyle \theta_p,\;\beta_p \;\gets\;
      \arg\min_{\theta_p,\beta_p}\;\sum_{(x,t)\sim\mathcal{D}_u}\,
        \Bigl\|\,r(x,t)\;-\;\Phi_p\bigl(x;\theta_p\bigr)\,\Psi_p\bigl(t;\beta_p\bigr)\Bigr\|_{2}^2$

    \State $r(x,t) \;\gets\;
      r(x,t)\;-\;\Phi_p\bigl(x;\theta_p\bigr)\,\Psi_p\bigl(t;\beta_p\bigr)$
  \EndWhile
\end{algorithmic}
\end{algorithm}

\paragraph{Architecture}

The proposed Neural-POD architecture comprises two distinct sub-networks, designated as the \textit{parameter} network and the \textit{basis} network, which are integrated into a composite framework termed \textit{Neural-POD}. This design strategically decouples the parameter-dependent representation from the basis function expansion, thereby facilitating a flexible approach to modeling complex datasets. Within the composite architecture, the Neural-POD network processes inputs by independently computing the outputs of the \textit{parameter} and \textit{basis} sub-networks. These outputs are then combined via element-wise multiplication, and the resulting product is summed over the appropriate dimensions to yield a single scalar output for each sample.


\paragraph{Loss Functions}
In the Neural-POD framework, the basis network is designed to generate a set of basis functions analogous to those used in finite element methods (FEM). These basis functions are assumed to satisfy the following:
\begin{enumerate}
    \item \textbf{Optimality:} They provide the optimal representation for the data with a given norm, e.g., $L^2$ optimality.
    \item \textbf{Orthogonality:} They are mutually orthogonal.
    \item \textbf{Boundary conditions:} They satisfy the prescribed boundary conditions of the original PDE systems.
\end{enumerate}
The \textit{optimality}, \textit{normalization} and  \textit{boundary conditions} can be enforced by involving them into the loss functions.
\begin{align}
    &\text{Optimality:   } &&L_{data} = \sum_{i=1}^N  \left\| \Phi(x_i)\Psi(t_i) -u_i \right\|_S,
    \\
    &\text{Boundary conditions:   }&& L_{bc} = \sum_{i=1}^N  \left( \Phi(x^{bc}_i) -u^{bc}_i \right)^2
    +
    \sum_{i=1}^N  \left( \frac{\partial\Phi(x^{bc}_i)}{\partial x} -\frac{\partial u^{bc}_i}{\partial x} \right)^2
\end{align}
To ensure a balanced optimization, reciprocal weights are computed from the initial loss contributions of the data fidelity and basis regularization terms. Consequently, the overall loss function employed during training is expressed as a weighted sum:
\begin{align}
L = w_{\text{data}} \cdot L_{\text{data}} + w_{\text{bc}} \cdot L_{\text{bc}},
\end{align}
where \(w_{\text{data}}\), \(w_{\text{basis}}\), and \(w_{\text{bc}}\) are the weights assigned to the data loss, basis regularization, and any additional boundary condition loss \(L_{\text{bc}}\), respectively. This balanced loss formulation promotes both an accurate data fit and a well-scaled basis function.
where \(w_{\text{data}}\), \(w_{\text{basis}}\), and \(w_{\text{bc}}\) are the weights assigned to the data loss, basis regularization, and any additional boundary condition loss \(L_{\text{bc}}\), respectively. These loss functions guarantee both an optimal data representation and a normalized basis function.

In Algorithm \ref{alg:neural-pod}, the temporal variation of the data is expressed as a neural network; however, training two neural network at a time is computationally expensive. One may consider to choose one of the neural network $\psi_k$ as a trainable function.

\subsection*{Neural-POD in Galerkin Projection Based Reduced Order Models }

We first introduce the Neural-POD framework in the context of Galerkin projection reduced-order models (Galerkin ROMs or POD-ROMs). Although the methodology is applicable to general POD-based ROM constructions for a broad class of parameterized PDEs, we use the one-dimensional viscous Burgers equation as a simple and illustrative example. Specifically, we consider the Burgers equation posed on a spatial domain $\Omega \subset \mathbb{R}$ over a finite time interval $[0,T]$. The solution is denoted by $u=u(x,t)$ for $(x,t)\in \Omega\times[0,T]$.The governing equation is equipped with Dirichlet boundary conditions and a prescribed initial condition:
\begin{align}
\left\{
\begin{aligned}
    & u_t - \nu u_{xx} + u u_x = 0\,, && (x,t) \in \Omega \times (0,T], \\
    & u(x,t) = 0\,, && (x,t) \in \partial\Omega \times (0,T], \\
    & u(x,0) = u_0(x)\,, && x \in \Omega \,.
\end{aligned}
\right.
\label{eq:burgers}
\end{align}

\paragraph{Traditional Projection-ROM}
To construct a ROM, we approximate the solution $u(x,t)$ by a linear combination of $r$ orthonormal basis functions $\{ \phi_i(x) \}_{i=1}^r$ that can be obtained either from POD or Neural-POD:
\begin{equation}
    u_r(x,t) = \sum_{i=1}^r a_i(t) \phi_i(x),
\end{equation}
where $a_i(t)$ are the time-dependent basis coefficients.
Substituting $u_r(x,t)$ into the Burgers’ equation~\eqref{eq:burgers} and applying Galerkin projection, we obtain the following:
\begin{equation}
    \left( \frac{\partial u_r}{\partial t} + u_r \frac{\partial u_r}{\partial x} - \nu \frac{\partial^2 u_r}{\partial x^2},\ \phi_k \right) = 0, \quad \text{for } k = 1,\dots,r,
\end{equation}
where $(\cdot,\cdot)$ denotes the standard $L^2$ inner product.
This yields the following dynamic system:
\begin{equation}
    \dot{a}_k(t) = - \sum_{i=1}^r \sum_{j=1}^r a_i(t)\, a_j(t)\, \left( \phi_i \frac{d\phi_j}{dx}, \phi_k \right) 
    + \nu \sum_{i=1}^r a_i(t)\, \left( \frac{d^2 \phi_i}{dx^2}, \phi_k \right), 
    \quad k = 1,\dots,r.
    \label{eq:galerkin_system}
\end{equation}
where
\begin{align}
    C_{ijk} &= \left( \phi_i \frac{d\phi_j}{dx}, \phi_k \right), \\
    D_{ik}  &= \left( \frac{\partial^2 \phi_i}{\partial x^2}, \phi_k \right).
\end{align}
With these definitions, equation~\eqref{eq:galerkin_system} becomes:
\begin{equation}
    \dot{a}_k(t) = - \sum_{i=1}^r \sum_{j=1}^r a_i(t)\, a_j(t)\, C_{ijk} 
    + \nu \sum_{i=1}^r a_i(t)\, D_{ik}, 
    \quad k = 1,\dots,r.
\end{equation}

\paragraph{Neural-POD-ROM}
In the Neural-POD-ROM framework, the ROM basis
$\{ \phi_i(x,\nu) \}_{i=1}^r$ is obtained from a pretrained Neural-POD model evaluated at a given PDE parameter setting—such as the viscosity $\nu$ in the Burgers equation—and a prescribed set of spatial grid points $\{x_j\}_{j=1}^N$, which may differ from those used during training, in contrast to classical snapshot-based POD, whose basis functions are fixed by the training snapshots and are therefore restricted to the viscosity and spatial resolution represented in the training data.

Once the Neural-POD modes are learned and fixed, the ROM solution can be expressed in the same form,
\begin{equation}
    u_r(x,t;\nu) = \sum_{i=1}^r a_i(t) \phi_i(x,\nu),
\end{equation}
and the governing equations are derived by substituting $u_r$ into the Burgers’ equation and applying Galerkin projection onto the Neural-POD basis.
This results in the same reduced dynamical system as in the traditional POD-ROM,
\begin{equation}
    \dot{a}_k(t) = - \sum_{i=1}^r \sum_{j=1}^r a_i(t)\, a_j(t)\, C_{ijk}
    + \nu \sum_{i=1}^r a_i(t)\, D_{ik},
    \quad k = 1,\dots,r,
\end{equation}
where the tensors $C_{ijk}$ and $D_{ik}$ are now computed using the Neural-POD modes. The projection-based ROM using POD and Neural-POD is specifically illustrated in Figure \ref{fig:npod-rom}.

\begin{remark}
A fundamental difference between Neural-POD-ROM and classical POD-ROM lies in the flexibility of the reduced basis: Neural-POD-ROM allows evaluation at new viscosity values and spatial resolutions, while classical POD-ROM remains tied to the parameter and grid of the training snapshots.
\end{remark}

\subsection*{Neural-POD in Deep Operator Networks}

Operator learning seeks to approximate mappings between infinite-dimensional function spaces and has become an effective framework for modeling PDE-governed systems.
Given an input function $v \in \mathcal{V}$ (e.g., forcing terms or initial conditions) and an output function $u \in \mathcal{U}$ (e.g., the PDE solution), the objective is to learn an operator
\begin{equation}
\mathcal{G} : \mathcal{V} \to \mathcal{U}, \qquad u = \mathcal{G}(v),
\end{equation}
from a dataset of paired input–output functions.
Deep Operator Network (DeepONet)~\cite{lu2019deeponet} is a representative neural operator architecture that realizes this mapping using a branch–trunk decomposition and has demonstrated strong performance in a variety of applications~\cite{wang2021learning,lu2022comprehensive}.

In vanilla DeepONet, the branch network encodes a discretized representation of the input function $v$, while the trunk network takes the output coordinates $\xi \in D'$ as input.
The operator output is expressed as
\begin{equation}
\mathcal{G}(v)(\xi)
=
\sum_{k=1}^{p} b_k(v)\, t_k(\xi) + b_0,
\end{equation}
where $\{b_k(v)\}$ and $\{t_k(\xi)\}$ denote the outputs of the branch and trunk networks, respectively.
In this formulation, the trunk network implicitly learns a basis for the output function space directly from data.

To improve interpretability and efficiency, POD--DeepONet~\cite{lu2022comprehensive} replaces the learned trunk basis with pre-obtained POD modes from the training data. As a result, after subtracting the mean field, $\phi_0(\xi)$, the output operator is represented as
\begin{equation}
\mathcal{G}(v)(\xi)
=
\sum_{k=1}^{p} b_k(v)\,\phi_k(\xi) + \phi_0(\xi),
\end{equation}
where $\{\phi_k\}_{k=1}^p$ are fixed POD basis functions and the coefficients $b_k(v)$ are learned by the branch network.

Following this framework, Neural-POD is applied within DeepONet by using pretrained Neural-POD modes in place of classical snapshot-based POD modes to construct the trunk representation.
In this case, the operator output is expressed as
\begin{equation}
\mathcal{G}(v)(\xi;\nu,\mathcal{X})
=
\sum_{k=1}^{p} b_k(v)\,\phi_k (\xi;\nu,\mathcal{X})
+
\phi_0 (\xi;\nu,\mathcal{X}),
\label{eq:neuralpod_deeponet}
\end{equation}
where $\phi_k (\cdot;\nu,\mathcal{X})$ denote the pretrained Neural-POD modes evaluated at viscosity $\nu$ and spatial grid $\mathcal{X}=\{x_i\}_{i=1}^{N}$, and $\phi_0 $ is the corresponding mean field.
The branch network outputs $b_k(v)$ represent the modal coefficients associated with these Neural-POD basis functions. The Neural-POD-DeepONet framework is illustrated in Figure \ref{fig:npod-deeponet}.

\section*{Acknowledgment}
We would like to thank the support of National Science Foundation (DMS-2533878, DMS-2053746, DMS-2134209, ECCS-2328241, CBET-2347401 and OAC-2311848), and U.S.~Department of Energy (DOE) Office of Science Advanced Scientific Computing Research program DE-SC0023161, the SciDAC LEADS Institute, and DOE–Fusion Energy Science, under grant number: DE-SC0024583.

\section*{Data availability}
Data used in this study are available at \url{https://github.com/chhmou/Neural-POD}
. The full dataset is available upon reasonable request to the authors of the associated paper.

\section*{Code availability}
The code used in this study, including scripts for data analysis in MATLAB (R2025b) and Python (v3.10), is available via GitHub at \url{https://github.com/chhmou/Neural-POD}. Python packages used for analysis include PyTorch, pandas, numpy, scikit-learn, matplotlib and statsmodels. Code annotations were generated using ChatGPT 4o (gpt-4o-2024-08-06).

\bibliography{ref,ref-rom}

@article{benner2015survey,
  title={A survey of projection-based model reduction methods for parametric dynamical systems},
  author={Benner, P. and Gugercin, S. and Willcox, K.},
  journal={SIAM Rev.},
  fjournal={SIAM Review},
  volume={57},
  number={4},
  pages={483--531},
  year={2015},
  publisher={SIAM}
}

@article{carlberg2011efficient,
  title={Efficient non-linear model reduction via a least-squares {P}etrov--{G}alerkin 
         projection and compressive tensor approximations},
  author={Carlberg, K. and Bou-Mosleh, C. and Farhat, C.},
  journal={Int. J. Num. Meth. Eng.},
  fulljournal={International Journal for Numerical Methods in Engineering},
  volume={86},
  number={2},
  pages={155--181},
  year={2011},
  publisher={Wiley Online Library}
}

@Book{HLB96,
  author = 	 {Holmes, P. and Lumley, J. L. and Berkooz, G.},
  title = 	 {Turbulence, Coherent Structures, Dynamical Systems 
                  and Symmetry},
  publisher = 	 {Cambridge},
  year = 	 {1996}
}

@article{iliescu2013variational,
  title={Variational multiscale proper orthogonal decomposition: {C}onvection-dominated convection-diffusion-reaction equations},
  author={Iliescu, T. and Wang, Z.},
  journal = {Math. Comput.},
  fjournal={Mathematics of Computation},
  volume={82},
  number={283},
  pages={1357--1378},
  year={2013}
}

@Article{iliescu2014are,
  author = 	 {Iliescu, T. and Wang, Z.},
  title = 	 {Are the Snapshot Difference Quotients Needed in the Proper Orthogonal Decomposition?},
  journal=       {SIAM J. Sci. Comput.},
  fjournal=      {SIAM Journal on Scientific Computing},
  year = 	 {2014},
  volume = 	 {36},
  number = 	 {3},
  pages = 	 {A1221--A1250}
}

@article {KV01,
    AUTHOR = {Kunisch, K. and Volkwein, S.},
     TITLE = {Galerkin proper orthogonal decomposition methods for parabolic
              problems},
   JOURNAL = {Numer. Math.},
  FJOURNAL = {Numerische Mathematik},
    VOLUME = {90},
      YEAR = {2001},
    NUMBER = {1},
     PAGES = {117--148},
      ISSN = {0029-599X},
     CODEN = {NUMMA7},
   MRCLASS = {65M60 (35A35 35K20)},
  MRNUMBER = {MR1868765 (2003g:65118)},
}

@article{kunisch2008proper,
  title={Proper orthogonal decomposition for optimality systems},
  author={Kunisch, K. and Volkwein, S.},
  journal={ESAIM: Math. Model. Numer. Anal.},
  fjournal={ESAIM: Mathematical Modelling and Numerical Analysis},
  volume={42},
  number={1},
  pages={1--23},
  year={2008},
  publisher={EDP Sciences}
}

@InCollection{Lay02c,
  author = 	 {Layton, W.J.},
  title = 	 {A Mathematical Introduction to Large Eddy
                  Simulation}, 
  booktitle = 	 {Computational Fluid Dynamics-Multiscale Methods},
  OPTcrossref =  {},
  OPTkey = 	 {},
  OPTpages = 	 {},
  publisher = {Von Karman Institute for Fluid Dynamics},
  OPTyear = 	 {2002},
  editor = 	 {H. Deconinck},
  OPTvolume = 	 {},
  OPTnumber = 	 {},
  OPTseries = 	 {},
  OPTtype = 	 {},
  OPTchapter = 	 {},
  address = 	 {Rhode-Saint-Gen\`ese, Belgium},
  OPTedition = 	 {},
  OPTmonth = 	 {},
  OPTnote = 	 {},
  OPTannote = 	 {}
}

@article{mou2020data,
  title={Data-Driven Correction Reduced Order Models for the Quasi-Geostrophic Equations: A Numerical Investigation},
  author={Mou, C. and Liu, H. and Wells, D. R. and Iliescu, T.},
  OPTjournal={arXiv preprint, \url{http://arxiv.org/abs/1908.05297}},
  journal={Int. J. Comput. Fluid Dyn.},
  pages={1--13},
  year={2020},
  publisher={Taylor \& Francis}
}

@book{quarteroni2015reduced,
  title={Reduced Basis Methods for Partial Differential Equations: An Introduction},
  author={Quarteroni, A. and Manzoni, A. and Negri, F.},
  volume={92},
  year={2015},
  publisher={Springer}
}

@article{rowley2004model,
  title={Model reduction for compressible flows using {POD} and {G}alerkin projection},
  author={Rowley, C. W. and Colonius, T. and Murray, R. M.},
  journal={Phys. D},
  fjournal={Physica D: Nonlinear Phenomena},
  volume={189},
  number={1},
  pages={115--129},
  year={2004},
  publisher={Elsevier}
}

@article{shadden2011lagrangian,
  title={Lagrangian coherent structures},
  author={Shadden, S. C.},
  journal={Transport and Mixing in Laminar Flows: From Microfluidics to Oceanic Currents},
  pages={59--89},
  year={2011},
  publisher={Wiley Online Library}
}

@article{taira2017modal,
  title={Modal Analysis of Fluid Flows: {A}n Overview},
  author={Taira, K. and Brunton, S. L. and Dawson, S. and Rowley, C. W. and Colonius, T. and McKeon, B. J. and Schmidt, O. T. and Gordeyev, S. and Theofilis, V. and Ukeiley, L. S.},
  journal={AIAA J.},
  fjournal={AIAA Journal},
  pages={4013--4041},
  year={2017},
  publisher={American Institute of Aeronautics and Astronautics}
}

@article{volkwein2013proper,
  title={Proper orthogonal decomposition: Theory and reduced-order modelling},
  author={Volkwein, S.},
  journal={Lecture Notes, University of Konstanz},
  note={\url{http://www.math.uni-konstanz.de/numerik/personen/volkwein/teaching/POD-Book.pdf}},
  year={2013}
}

@article{li2020fourier,
  title={Fourier neural operator for parametric partial differential equations},
  author={Li, Zongyi and Kovachki, Nikola and Azizzadenesheli, Kamyar and Liu, Burigede and Bhattacharya, Kaushik and Stuart, Andrew and Anandkumar, Anima},
  journal={arXiv preprint arXiv:2010.08895},
  year={2020}
}

@article{wang2021learning,
  title={Learning the solution operator of parametric partial differential equations with physics-informed DeepONets},
  author={Wang, Sifan and Wang, Hanwen and Perdikaris, Paris},
  journal={Science advances},
  volume={7},
  number={40},
  pages={eabi8605},
  year={2021},
  publisher={American Association for the Advancement of Science}
}

@article{lucia2004reduced,
  title={Reduced-order modeling: new approaches for computational physics},
  author={Lucia, David J and Beran, Philip S and Silva, Walter A},
  journal={Progress in aerospace sciences},
  volume={40},
  number={1-2},
  pages={51--117},
  year={2004},
  publisher={Elsevier}
}

@article{yu2019non,
  title={Non-intrusive reduced-order modeling for fluid problems: A brief review},
  author={Yu, Jian and Yan, Chao and Guo, Mengwu},
  journal={Proceedings of the Institution of Mechanical Engineers, Part G: Journal of Aerospace Engineering},
  volume={233},
  number={16},
  pages={5896--5912},
  year={2019},
  publisher={SAGE Publications Sage UK: London, England}
}

@article{mou2021data,
  title={Data-driven variational multiscale reduced order models},
  author={Mou, Changhong and Koc, Birgul and San, Omer and Rebholz, Leo G and Iliescu, Traian},
  journal={Computer Methods in Applied Mechanics and Engineering},
  volume={373},
  pages={113470},
  year={2021},
  publisher={Elsevier}
}

@article{lu2019deeponet,
  title={Deeponet: Learning nonlinear operators for identifying differential equations based on the universal approximation theorem of operators},
  author={Lu, Lu and Jin, Pengzhan and Karniadakis, George Em},
  journal={arXiv preprint arXiv:1910.03193},
  year={2019}
}

@article{lu2022comprehensive,
  title={A comprehensive and fair comparison of two neural operators (with practical extensions) based on fair data},
  author={Lu, Lu and Meng, Xuhui and Cai, Shengze and Mao, Zhiping and Goswami, Somdatta and Zhang, Zhongqiang and Karniadakis, George Em},
  journal={Computer Methods in Applied Mechanics and Engineering},
  volume={393},
  pages={114778},
  year={2022},
  publisher={Elsevier}
}

@article{yosinski2014transferable,
  title={How transferable are features in deep neural networks?},
  author={Yosinski, Jason and Clune, Jeff and Bengio, Yoshua and Lipson, Hod},
  journal={Advances in neural information processing systems},
  volume={27},
  year={2014}
}

@inproceedings{deng2021deeplight,
  title={Deeplight: Deep lightweight feature interactions for accelerating ctr predictions in ad serving},
  author={Deng, Wei and Pan, Junwei and Zhou, Tian and Kong, Deguang and Flores, Aaron and Lin, Guang},
  booktitle={Proceedings of the 14th ACM international conference on Web search and data mining},
  pages={922--930},
  year={2021}
}

@article{mendez2019multi,
  title={Multi-scale proper orthogonal decomposition of complex fluid flows},
  author={Mendez, Miguel A and Balabane, Mikha{\"e}l and Buchlin, J-M},
  journal={Journal of Fluid Mechanics},
  volume={870},
  pages={988--1036},
  year={2019},
  publisher={Cambridge University Press}
}

@article{zahr2015progressive,
  title={Progressive construction of a parametric reduced-order model for PDE-constrained optimization},
  author={Zahr, Matthew J and Farhat, Charbel},
  journal={International Journal for Numerical Methods in Engineering},
  volume={102},
  number={5},
  pages={1111--1135},
  year={2015},
  publisher={Wiley Online Library}
}

@article{winovich2025active,
  title={Active operator learning with predictive uncertainty quantification for partial differential equations},
  author={Winovich, Nick and Daneker, Mitchell and Lu, Lu and Lin, Guang},
  journal={arXiv preprint arXiv:2503.03178},
  year={2025}
}

@article{lu2025evolutionary,
  title={An Evolutionary Multi-objective Optimization for Replica-Exchange-based Physics-informed Operator Learning Network},
  author={Lu, Binghang and Mou, Changhong and Lin, Guang},
  journal={arXiv preprint arXiv:2509.00663},
  year={2025}
}

@article{berkooz1993proper,
  title={The proper orthogonal decomposition in the analysis of turbulent flows},
  author={Berkooz, Gal and Holmes, Philip and Lumley, John L},
  journal={Annual review of fluid mechanics},
  volume={25},
  number={1},
  pages={539--575},
  year={1993},
  publisher={Annual Reviews 4139 El Camino Way, PO Box 10139, Palo Alto, CA 94303-0139, USA}
}

@article{chatterjee2000introduction,
  title={An introduction to the proper orthogonal decomposition},
  author={Chatterjee, Anindya},
  journal={Current science},
  pages={808--817},
  year={2000},
  publisher={JSTOR}
}

@article{gubisch2017proper,
  title={Proper orthogonal decomposition for linear-quadratic optimal control},
  author={Gubisch, Martin and Volkwein, Stefan},
  journal={Model reduction and approximation: theory and algorithms},
  volume={15},
  number={1},
  year={2017},
  publisher={SIAM Philadelphia}
}

@article{berkooz1992observations,
  title={Observations on the proper orthogonal decomposition},
  author={Berkooz, Gal},
  journal={Studies in Turbulence},
  pages={229--247},
  year={1992},
  publisher={Springer}
}

@article{mifsud2016variable,
  title={A variable-fidelity aerodynamic model using proper orthogonal decomposition},
  author={Mifsud, MJ and MacManus, David G and Shaw, ST},
  journal={International Journal for Numerical Methods in Fluids},
  volume={82},
  number={10},
  pages={646--663},
  year={2016},
  publisher={Wiley Online Library}
}

@article{roderick2014proper,
  title={Proper orthogonal decompositions in multifidelity uncertainty quantification of complex simulation models},
  author={Roderick, Oleg and Anitescu, Mihai and Peet, Yulia},
  journal={International Journal of Computer Mathematics},
  volume={91},
  number={4},
  pages={748--769},
  year={2014},
  publisher={Taylor \& Francis}
}

@article{girfoglio2021pod,
  title={A POD-Galerkin reduced order model for a LES filtering approach},
  author={Girfoglio, Michele and Quaini, Annalisa and Rozza, Gianluigi},
  journal={Journal of Computational Physics},
  volume={436},
  pages={110260},
  year={2021},
  publisher={Elsevier}
}

@article{tsai2025time,
  title={A time-relaxation reduced order model for the turbulent channel flow},
  author={Tsai, Ping-Hsuan and Fischer, Paul and Iliescu, Traian},
  journal={Journal of Computational Physics},
  volume={521},
  pages={113563},
  year={2025},
  publisher={Elsevier}
}

@article{koc2023uniform,
  title={Uniform bounds with difference quotients for proper orthogonal decomposition reduced order models of the Burgers equation},
  author={Koc, Birgul and Chacon Rebollo, Tomas and Rubino, Samuele},
  journal={Journal of Scientific Computing},
  volume={95},
  number={2},
  pages={43},
  year={2023},
  publisher={Springer}
}

@article{koc2022verifiability,
  title={Verifiability of the data-driven variational multiscale reduced order model},
  author={Koc, Birgul and Mou, Changhong and Liu, Honghu and Wang, Zhu and Rozza, Gianluigi and Iliescu, Traian},
  journal={Journal of Scientific Computing},
  volume={93},
  number={2},
  pages={54},
  year={2022},
  publisher={Springer}
}

@article{mou2023energy,
  title={An energy-based lengthscale for reduced order models of turbulent flows},
  author={Mou, Changhong and Merzari, Elia and San, Omer and Iliescu, Traian},
  journal={Nuclear Engineering and Design},
  volume={412},
  pages={112454},
  year={2023},
  publisher={Elsevier}
}

@article{lu2025mopinnenkf,
  title={MoPINNEnKF: Iterative Model Inference using generic-PINN-based ensemble Kalman filter},
  author={Lu, Binghang and Mou, Changhong and Lin, Guang},
  journal={arXiv preprint arXiv:2506.00731},
  year={2025}
}

@article{popov2021multifidelity,
  title={A multifidelity ensemble Kalman filter with reduced order control variates},
  author={Popov, Andrey A and Mou, Changhong and Sandu, Adrian and Iliescu, Traian},
  journal={SIAM Journal on Scientific Computing},
  volume={43},
  number={2},
  pages={A1134--A1162},
  year={2021},
  publisher={SIAM}
}

@article{cai2021physics,
  title={Physics-informed neural networks (PINNs) for fluid mechanics: A review},
  author={Cai, Shengze and Mao, Zhiping and Wang, Zhicheng and Yin, Minglang and Karniadakis, George Em},
  journal={Acta Mechanica Sinica},
  volume={37},
  number={12},
  pages={1727--1738},
  year={2021},
  publisher={Springer}
}

@article{bilionis2013multi,
  title={Multi-output separable Gaussian process: Towards an efficient, fully Bayesian paradigm for uncertainty quantification},
  author={Bilionis, Ilias and Zabaras, Nicholas and Konomi, Bledar A and Lin, Guang},
  journal={Journal of Computational Physics},
  volume={241},
  pages={212--239},
  year={2013},
  publisher={Elsevier}
}

@article{bright2013compressive,
  title={Compressive sensing based machine learning strategy for characterizing the flow around a cylinder with limited pressure measurements},
  author={Bright, Ido and Lin, Guang and Kutz, J Nathan},
  journal={Physics of Fluids},
  volume={25},
  number={12},
  year={2013},
  publisher={AIP Publishing}
}

@article{lin2009efficient,
  title={An efficient, high-order probabilistic collocation method on sparse grids for three-dimensional flow and solute transport in randomly heterogeneous porous media},
  author={Lin, Guang and Tartakovsky, Alexandre M},
  journal={Advances in Water Resources},
  volume={32},
  number={5},
  pages={712--722},
  year={2009},
  publisher={Elsevier}
}

@incollection{xiu2025computational,
  title={COMPUTATIONAL FRAMEWORK FOR REAL-TIME DIGITAL TWINS},
  author={Xiu, Dongbin and Tartakovsky, Daniel M},
  booktitle={Thermopedia},
  year={2025},
  publisher={Begel House Inc.}
}

@article{jones2020characterising,
  title={Characterising the Digital Twin: A systematic literature review},
  author={Jones, David and Snider, Chris and Nassehi, Aydin and Yon, Jason and Hicks, Ben},
  journal={CIRP journal of manufacturing science and technology},
  volume={29},
  pages={36--52},
  year={2020},
  publisher={Elsevier}
}

@article{shukla2024deep,
  title={Deep neural operators as accurate surrogates for shape optimization},
  author={Shukla, Khemraj and Oommen, Vivek and Peyvan, Ahmad and Penwarden, Michael and Plewacki, Nicholas and Bravo, Luis and Ghoshal, Anindya and Kirby, Robert M and Karniadakis, George Em},
  journal={Engineering Applications of Artificial Intelligence},
  volume={129},
  pages={107615},
  year={2024},
  publisher={Elsevier}
}

@article{pearson1901liii,
  title={LIII. On lines and planes of closest fit to systems of points in space},
  author={Pearson, Karl},
  journal={The London, Edinburgh, and Dublin philosophical magazine and journal of science},
  volume={2},
  number={11},
  pages={559--572},
  year={1901},
  publisher={Taylor \& Francis}
}

@article{hotelling1933analysis,
  title={Analysis of a complex of statistical variables into principal components.},
  author={Hotelling, Harold},
  journal={Journal of educational psychology},
  volume={24},
  number={6},
  pages={417},
  year={1933},
  publisher={Warwick \& York}
}

@article{lumley1967similarity,
  title={Similarity and the turbulent energy spectrum},
  author={Lumley, JL},
  journal={The physics of fluids},
  volume={10},
  number={4},
  pages={855--858},
  year={1967},
  publisher={AIP Publishing}
}

\clearpage
\section*{Extended Data Figures and Tables}

\captionsetup[figure]{name=Extended Data Fig.}
\captionsetup[table]{name=Extended Data Table}

\setcounter{figure}{0}
\setcounter{table}{0}



\begin{figure}[H]
    \centering
    \includegraphics[width=\linewidth]{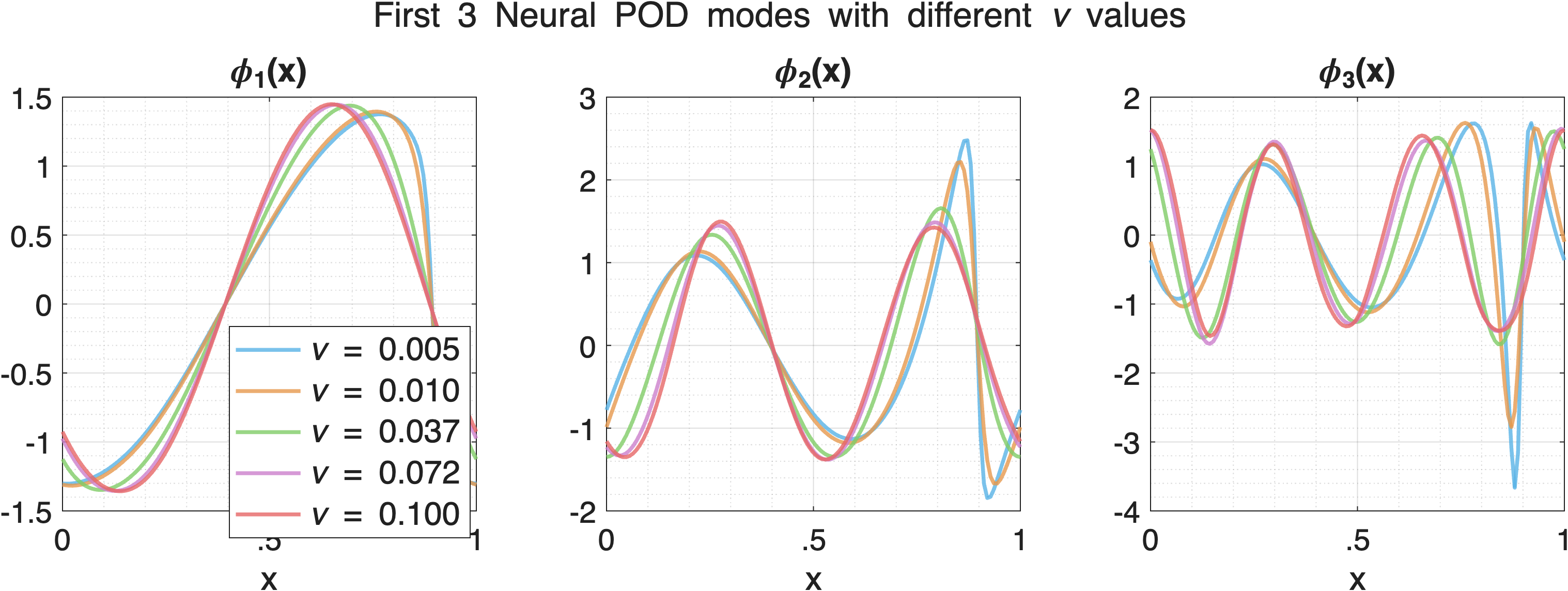}
    \caption{First three Neural-POD modes for different viscosity values
    }
    \label{fig:traditional-POD}
\end{figure}

\begin{figure}[H]
    \centering
    \includegraphics[width=\linewidth]{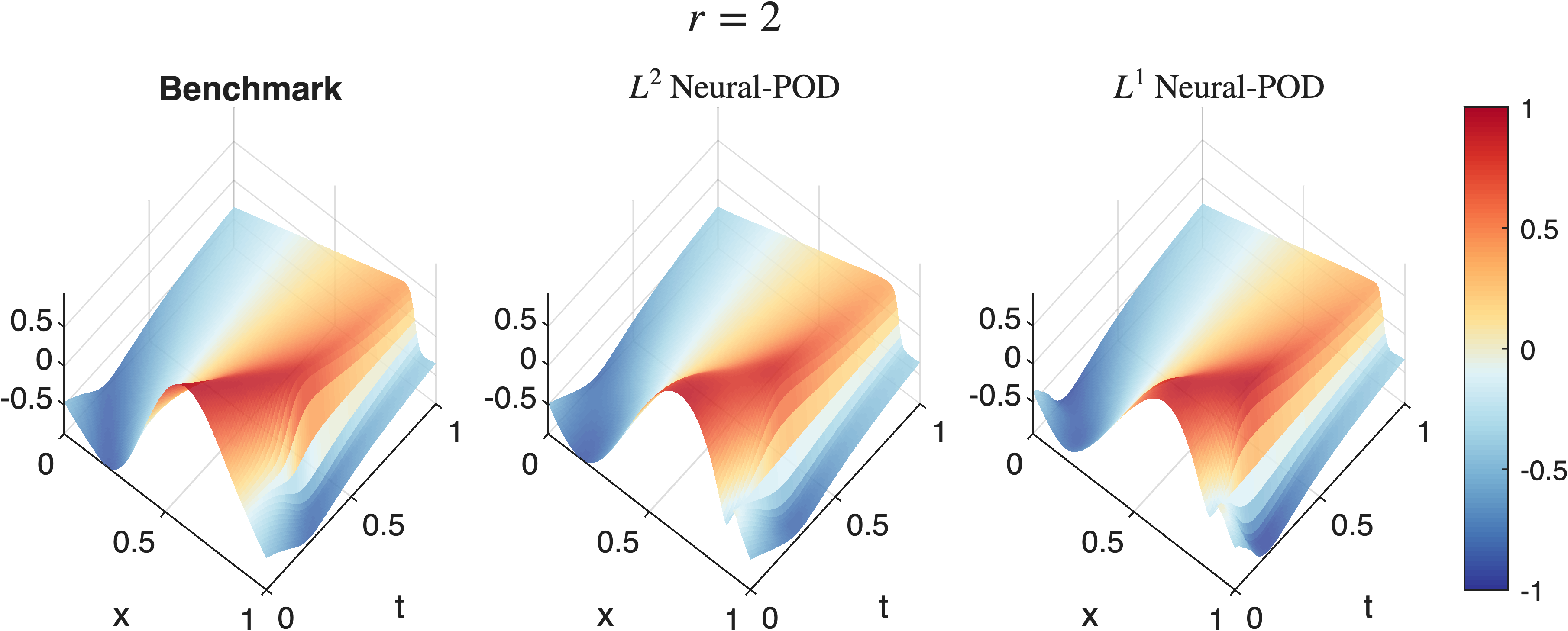}
    \includegraphics[width=\linewidth]{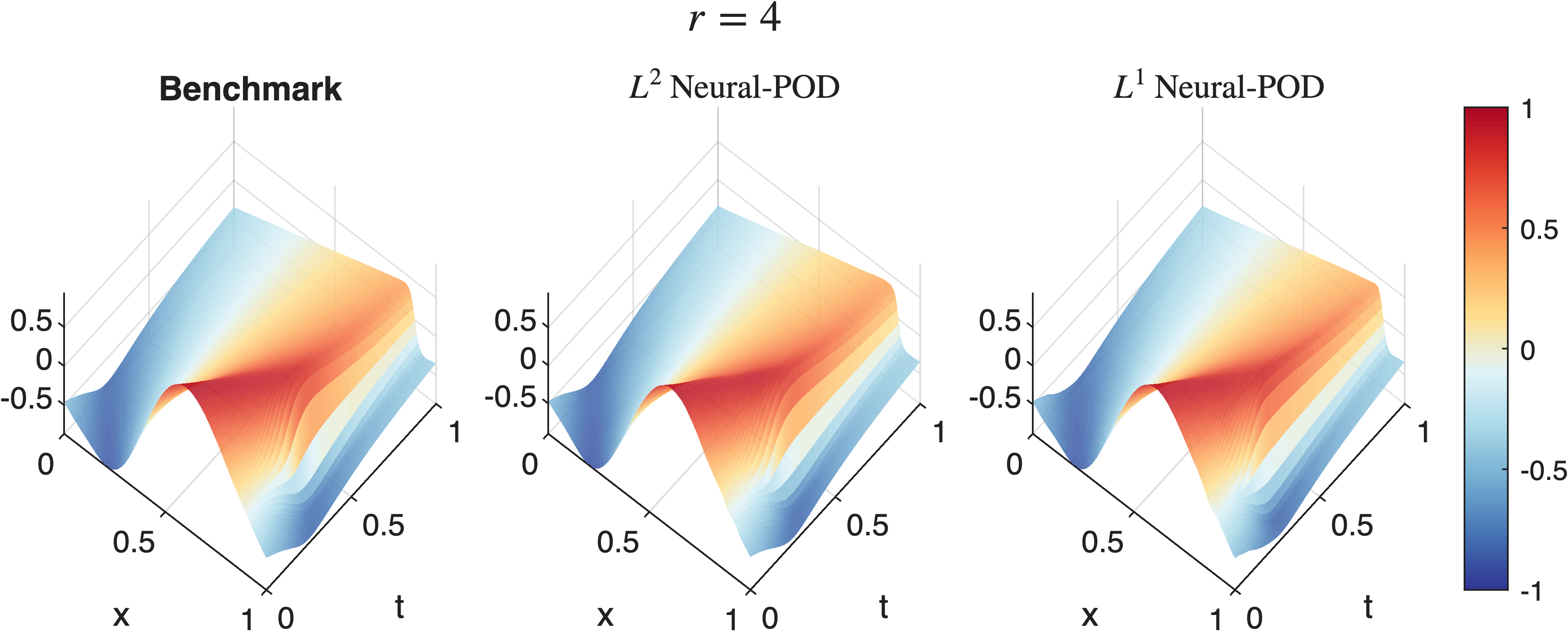}
    \caption{Reconstruction of spatiotemporal fields for the one-dimensional Burgers’ equation at an unseen viscosity parameter using Neural-POD with different latent dimensions (2 and 4).
    }
    \label{fig:neural-POD-recon}
\end{figure}



\begin{figure}[H]
    \centering
    \begin{subfigure}[b]{\linewidth}
        \centering
        \includegraphics[width=\linewidth]{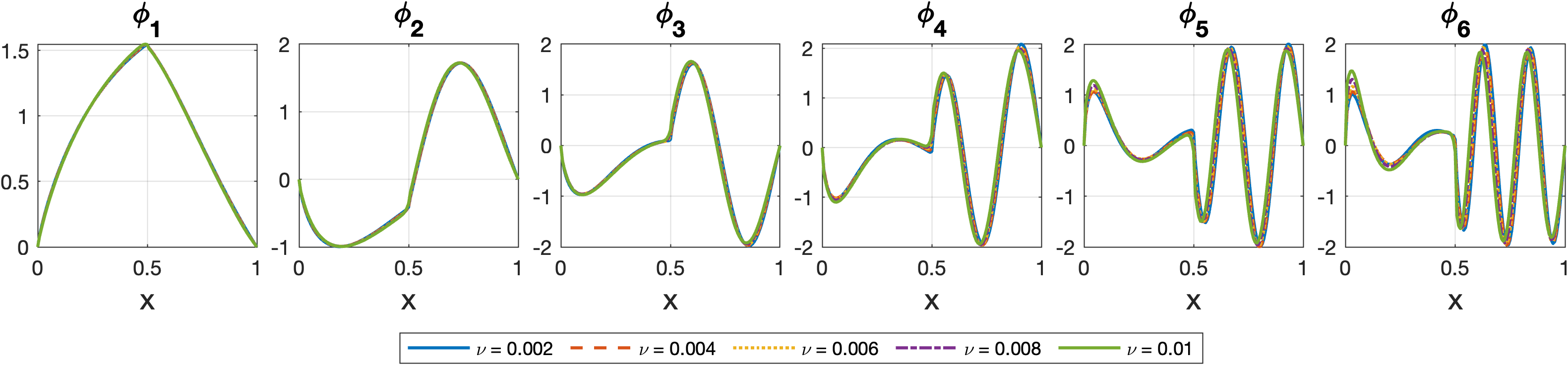}
        \caption{POD results for the 1D Burgers equation.}
        \label{fig:pod-burgers-l2}
    \end{subfigure}
    \begin{subfigure}[b]{\linewidth}
        \centering
        \includegraphics[width=\linewidth]{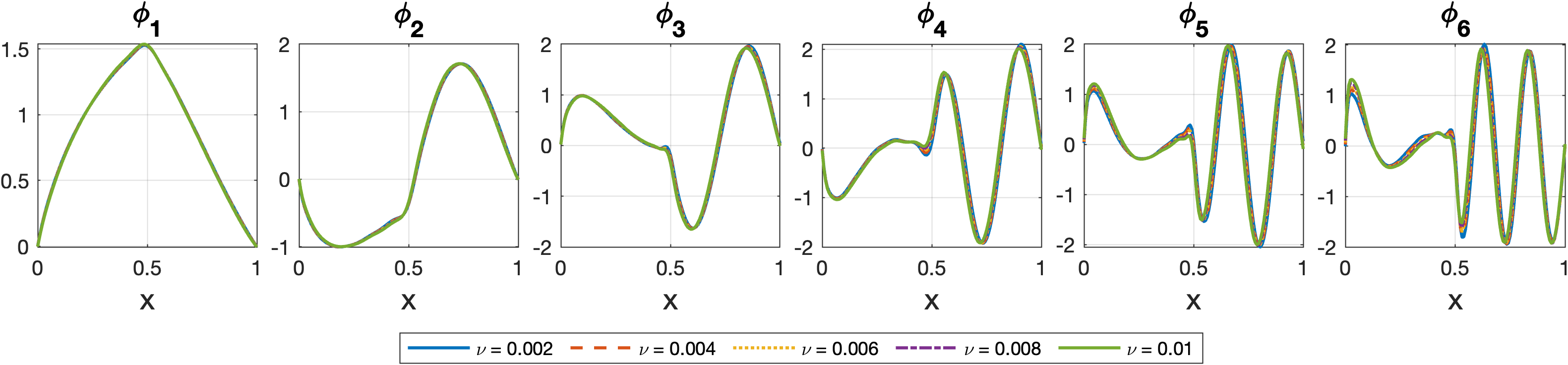}
        \caption{Neural-POD results for the 1D Burgers equation ($L^2$ optimality).}
        \label{fig:fun-neural-pod-burgers-l2}
    \end{subfigure}
    \begin{subfigure}[b]{\linewidth}
        \centering
        \includegraphics[width=\linewidth]{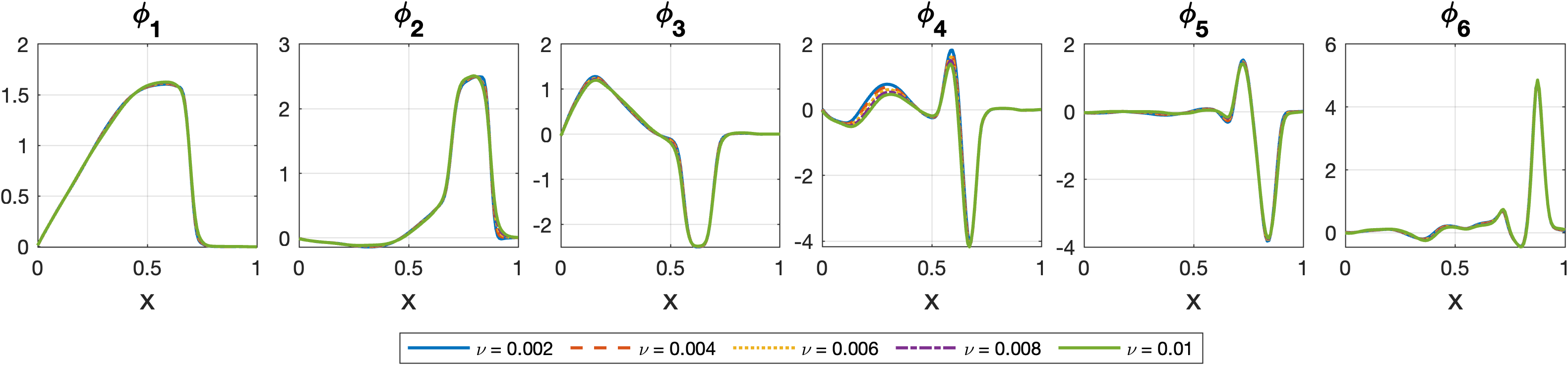}
        \caption{Neural-POD results for the 1D Burgers equation ($L^1$ optimality).}
        \label{fig:fun-neural-pod-burgers-l1}
    \end{subfigure}
    \caption{Comparison of POD and Neural-POD on the 1D Burgers equation for different viscosity values: 
    (a) classical POD , 
    (b) Neural-POD trained with $L^2$ loss, 
    (c) Neural-POD trained with $L^1$ loss.}
    \label{fig:pod-neural-burgers-comparison}
\end{figure}


\begin{table}[h]
\caption{
$L^2$ and $L^1$ error comparison between Neural-POD-DeepONet and POD-DeepONet models
trained on 64 spatial points with 100{,}000 training iterations.
}
\label{tab:compare_pod_models-2}
\begin{tabular*}{\textwidth}{@{\extracolsep\fill}lcc}
\toprule
Metric & Neural-POD-DeepONet & POD-DeepONet \\
\midrule
Training Loss       & \boldmath{$3.47 \times 10^{-3}$} & $4.10 \times 10^{-3}$ \\
Testing $L_2$ Error & \boldmath{$3.40 \times 10^{-3}$} & $4.37 \times 10^{-3}$ \\
Testing $L_1$ Error & \boldmath{$3.16 \times 10^{-2}$} & $3.59 \times 10^{-2}$ \\
\botrule
\end{tabular*}
\end{table}

\begin{figure}[H]
    \centering
    \begin{subfigure}[b]{\linewidth}
        \centering
        \includegraphics[width=\linewidth]{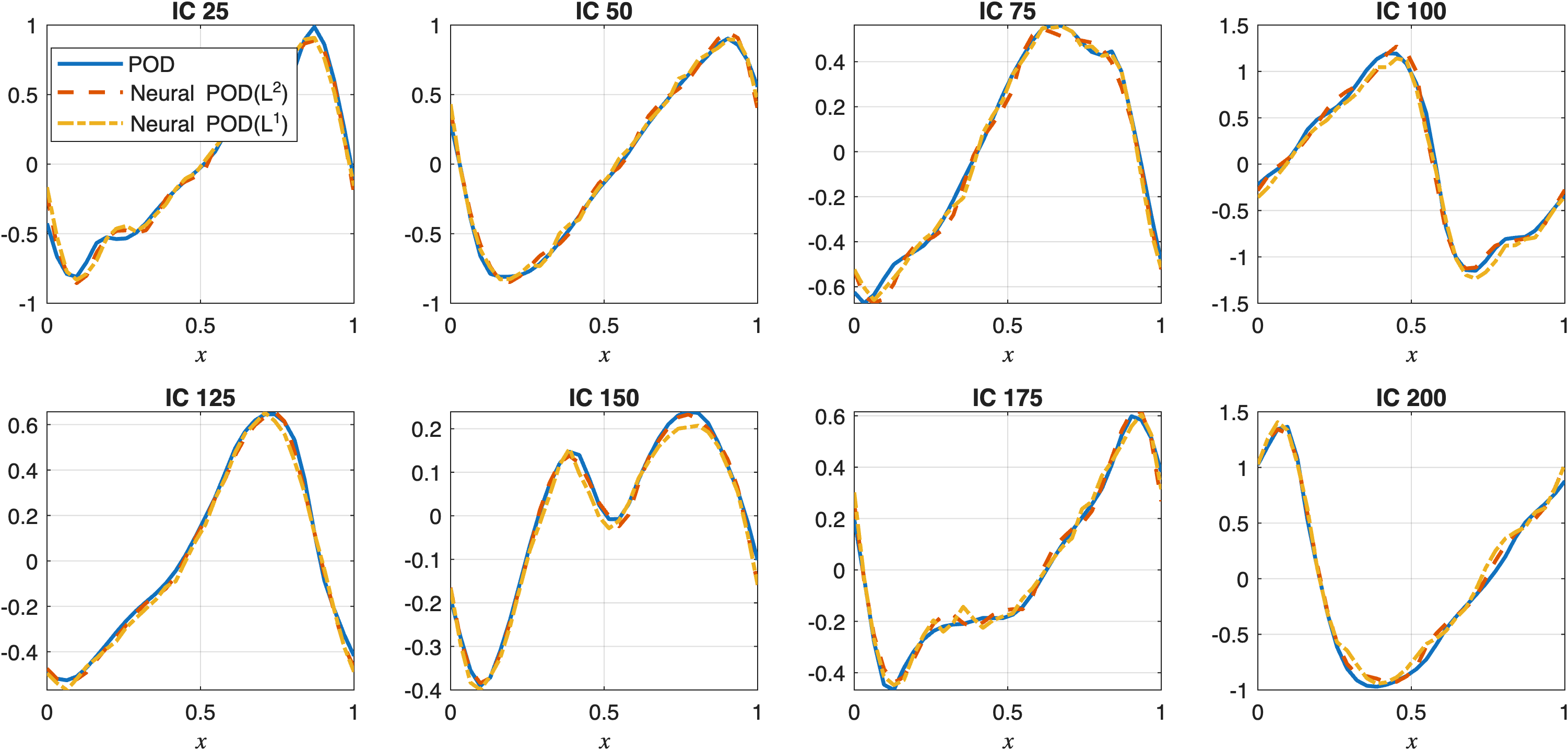}
        \caption{Training resolution $N=32$}
    \end{subfigure}
    
    \begin{subfigure}[b]{\linewidth}
        \centering
        \includegraphics[width=\linewidth]{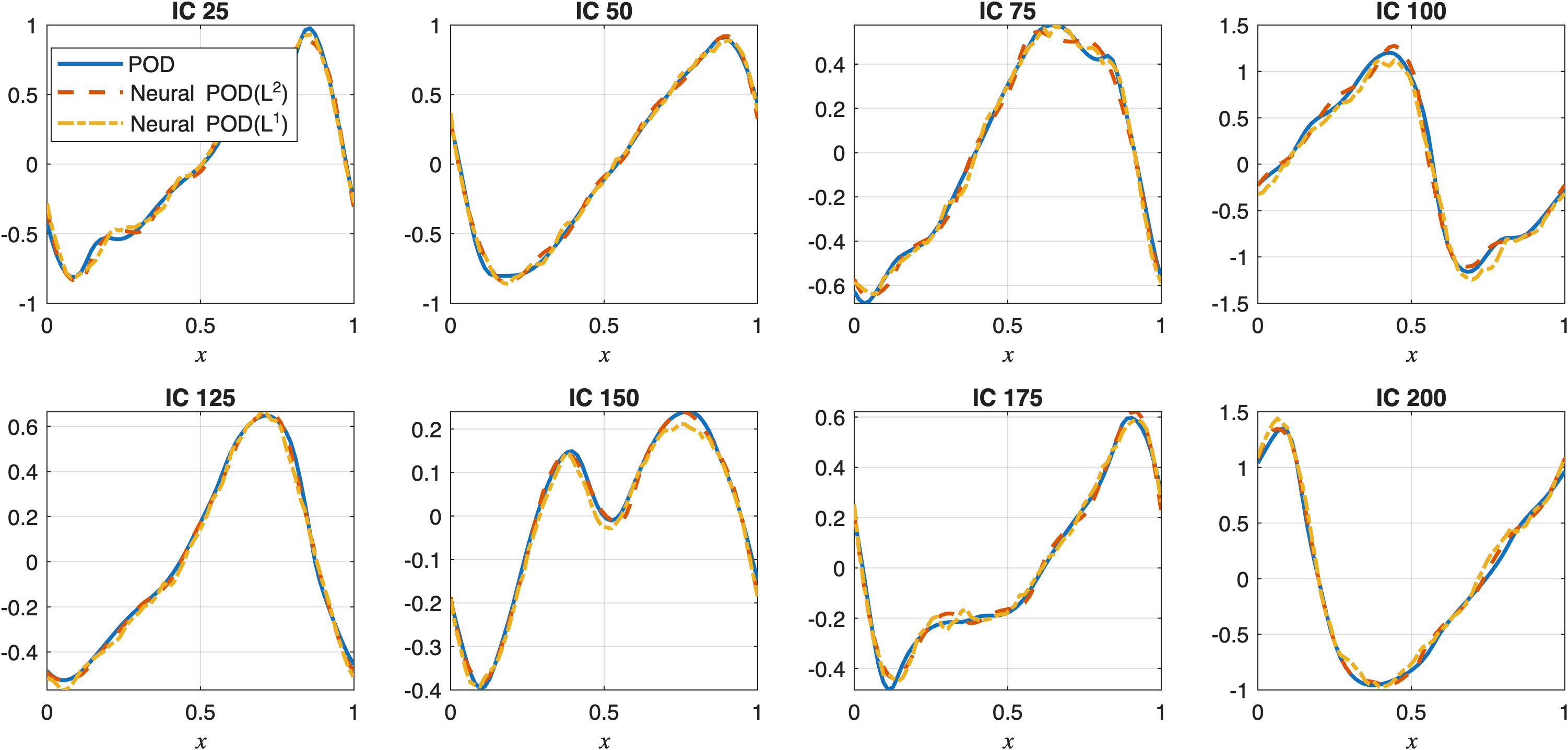}
        \caption{Training resolution $N=64$}
    \end{subfigure}
    
    \begin{subfigure}[b]{\linewidth}
        \centering
        \includegraphics[width=\linewidth]{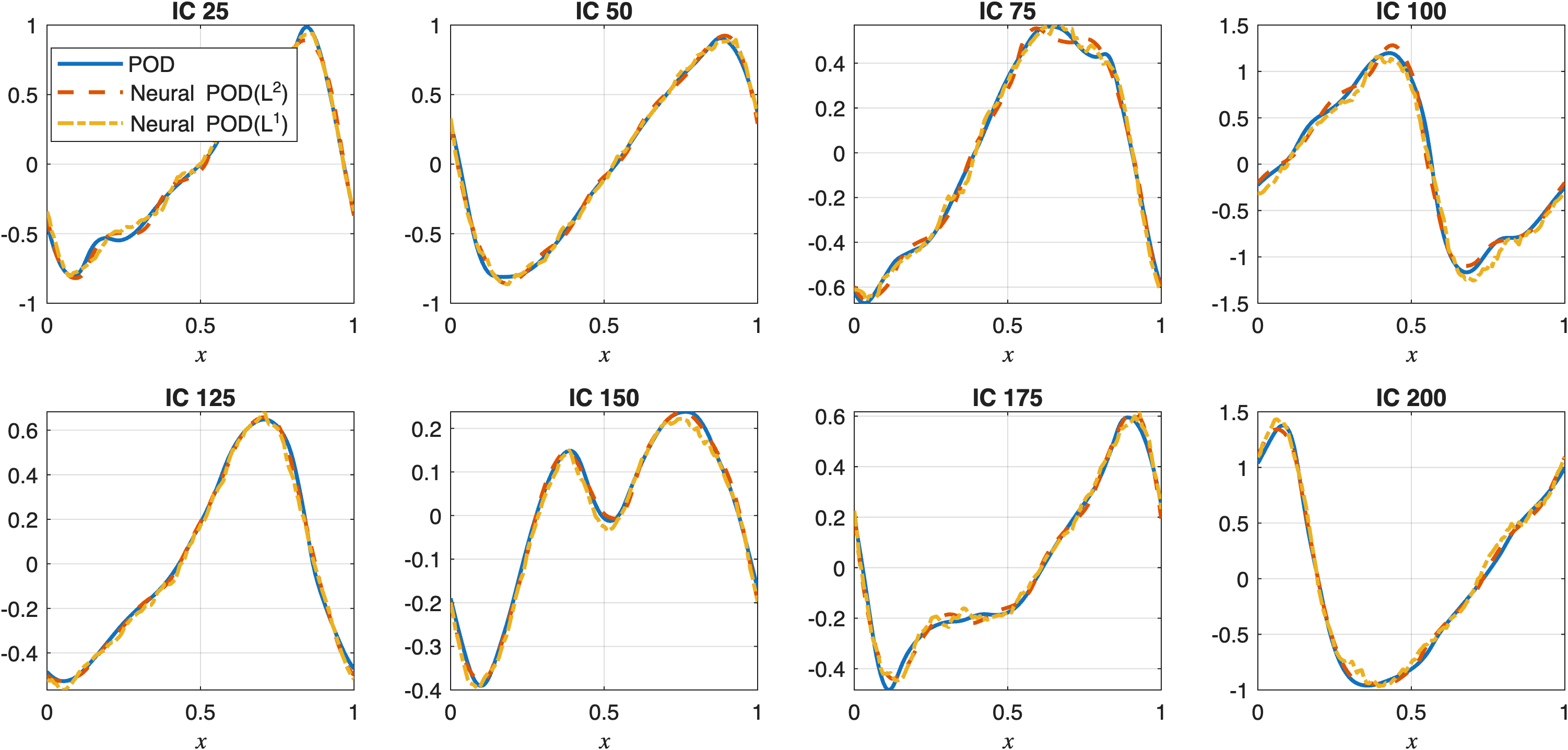}
        \caption{Training resolution $N=128$}
    \end{subfigure}
 \caption{\textbf{Resolution-independent} Neural-POD in DeepONet: A comparison of $L_1$ and $L_2$ errors for Neural-POD-DeepOnet versus POD-DeepOnet for different initial conditions and training resolutions.}   
    \label{fig:npod-deeponet-error}
\end{figure}



\end{document}